\documentclass[twocolumn,trackchanges]{aastex701}
\usepackage{amsmath}
\usepackage{makecell}
\usepackage{booktabs}

\begin{document}

\title{Unveiling Accretion-Disk Reverberation Beneath BLR Contamination with \texttt{TILDE}}

\author[orcid=0009-0009-5870-4266]{Sungho Jung}
\affiliation{Department of Physics \& Astronomy, Seoul National University, Seoul 08826, Republic of Korea}
\email{eric2912@snu.ac.kr}  

\author[orcid=0000-0002-8055-5465]{Jong-Hak Woo} 
\affiliation{Department of Physics \& Astronomy, Seoul National University, Seoul 08826, Republic of Korea}
\affiliation{SNU Astronomy Research Center, Seoul National University, Seoul 08826, Republic of Korea}
\email[show]{jhwoo@snu.ac.kr}

\begin{abstract}
We present a Bayesian framework, Two-stage Inference for Light curve DEcomposition (\texttt{TILDE}), for mapping the accretion disk and the broad line region (BLR) from active galactic nuclei (AGN) multi-band light curves. By inferring delayed BLR contamination from the underlying disk variability, \texttt{TILDE} provides the delay measurements of the disk as well as the BLR. We validate \texttt{TILDE} using synthetic data, finding that \texttt{TILDE} recovers the known time delays of the disk and BLR. By applying \texttt{TILDE} to the Seoul National University AGN Monitoring Project sample, we find that the inferred BLR light curves match the observed H$\beta$ emission light curves, providing consistent  BLR lags compared to the reported H$\beta$ lags. Using the lag spectrum, effectively corrected for the BLR contamination, we find that the disk size is broadly consistent with the standard thin-disk model, suggesting that BLR contamination can significantly bias the disk size derived from the cross-correlation analysis of the broad-band photometry. These results demonstrate that the new inference method is a powerful tool for determining both disk and BLR sizes from broad-band photometry in the era of time-domain surveys.
\end{abstract}


\section{Introduction}

Active galactic nuclei (AGN) exhibit strong stochastic variability over timescales from hours to years  \citep[see a recent review by][]{Paolillo2025}. Originating from sub-parsec regions, the delays of this variability between different emission components provide one of the powerful probes of the AGN structure. Reverberation mapping (RM) uses these delays to constrain the structure of the central engine \citep[see a review by][]{Cackett2021}, connecting nuclear accretion physics to black hole growth and host-galaxy coevolution. 

Spectroscopic RM determines the lag between the variable UV/optical continuum and the response of broad emission lines, providing broad line region (BLR) sizes and virial black hole mass estimates. Intensive monitoring campaigns have extended such reverberation studies across broader AGN populations \citep[e.g.,][]{Shen2015,Du2018,U2022,Woo2024}. However, spectroscopic RM requires time-consuming monitoring, leaving robust measurements limited to relatively small samples and restricted ranges of luminosity and redshift \citep[see a recent compilation by ][]{WangWoo2024}.

AGN variability also probes the accretion disk through continuum RM, which measures wavelength-dependent inter-band lags. In the standard geometrically thin, optically thick disk model \citep{Shakura1973}, shorter-wavelength emission arises from smaller radii; hence, the lag spectrum is expected to follow $\tau \propto \lambda^{4/3}$ \citep[e.g.,][]{Cackett2007,Fausnaugh2016}. Reported continuum lags generally follow this trend; however, inferred disk sizes are often factors of $\sim$3--5 larger than standard disk predictions \citep[e.g.,][]{Jha2022,Guo2022b,Mandal2025}, while broadly consistent with the size estimates from microlensing \citep{Morgan2010,Blackburne2011,Mosquera2013}. The problem of oversized disks may reflect additional reprocessing components \citep[e.g.,][]{Korista2001,Korista2019,Lawther2018,Netzer2020,Netzer2022,Lewin2025}, intrinsic departures from a homogeneous, geometrically thin disk, such as local temperature inhomogeneities or vertically structured disks \citep[e.g.,][]{DexterAgol2011,Starkey2023}, viewing and irradiation geometry \citep[e.g.,][]{Kammoun2019,Kammoun2021a,Kammoun2021b}, or systematic biases in the measured broad-band lags \citep[e.g.,][]{Tie2018,Chelouche2019,Jiang2024}.

A key complication in disk mapping is that broad-band photometry rarely isolates a pure accretion-disk continuum, as broad-band filters include emission from the BLR \citep[e.g.,][]{Lawther2018,Guo2022,Netzer2022}. The observed flux in each band is therefore a mixture of prompt disk variability and delayed BLR-related reprocessing. Inter-band lags measured from such light curves are likely to be contaminated by the delayed BLR emission depending on the relative weight of BLR emission observed in the filter. Thus, without correcting for the broad line contamination, inter-band lags and the correspondingly estimated disk sizes can be overestimated \citep[e.g.,][]{Peterson2014,CheloucheDaniel2012,Read2020}.

Broad-band photometric RM becomes increasingly important as wide-field surveys and dedicated monitoring programs can cover much larger AGN than is practical with spectroscopic RM \citep[e.g.,][]{Haas2011,PozoNunez2012,Jiang2016,PozoNunez2023}. This will become especially powerful in the era of large monitoring surveys, e.g., the Vera C. Rubin Observatory Legacy Survey of Space and Time \citep[LSST;][]{LSSTScienceCollaboration2009}. Nevertheless, extracting BLR lag from broad-band light curves remains challenging because the line contribution is typically modest compared to the continuum. Existing approaches rely on narrow/intermediate bands \citep[][]{Haas2011,Jiang2016,Woo2019NGC4395,MaWu2023} or on empirical continuum subtraction from broad-band light curves \citep[][]{Ma2023,Ma2024,Gu2024}. These approaches can become unstable when the line signal is weak, noisy, or blended with other emission components.

Motivated by aforementioned issues, we develop \texttt{TILDE} (Two-stage Inference for Light curve DEcomposition), a Bayesian framework that decomposes multi-band AGN light curves into variable accretion-disk and BLR components. By jointly modeling these components, \texttt{TILDE} aims to properly determine the disk and the BLR sizes from broad-band photometry. We describe the synthetic and the observed data used for validation in Section~\ref{Data}, and present \texttt{TILDE} in Section~\ref{sec:method}. We present the validation results in Section~\ref{Results}, discuss the implications and limitations in Section~\ref{sec:discussion}, and summarize our conclusions in Section~\ref{sec:summary}.

\begin{figure*}[ht!]
\centering
\includegraphics[width=\textwidth]{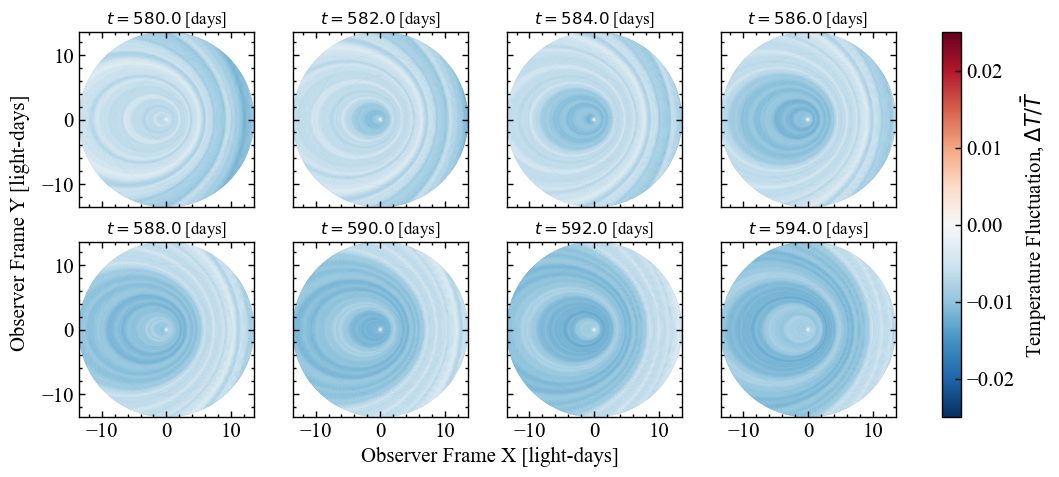}
\caption{
Snapshots of the fractional accretion-disk temperature fluctuation, $\Delta T/T$, induced by a variable central ionizing source in the lamp-post approximation. Each panel shows the disk in the observer-frame projected coordinates at a different epoch. The disk is inclined by $45^\circ$ with respect to LOS, with the observer located toward the left side of the image. The spiral-like temperature patterns reflect the light-travel-time delay across the inclined disk.
}
\label{fig:disk_temperature_snapshots}
\end{figure*}

\section{Test Datasets}
\label{Data}
We use two complementary datasets to validate \texttt{TILDE}. First, we generate synthetic broad-band light curves using a newly developed SED (Spectral Energy Distribution) simulator (\S\ref{sec:mock_data}). Second, we use the broad-band light curves from the Seoul National University AGN Monitoring Program \citep[SAMP;][]{Woo2024}, for which spectroscopic H$\beta$ light curves are also available (\S\ref{subsec:samp_data_details}). 

The synthetic data test whether \texttt{TILDE} can properly separate the disk and BLR light curves, while the SAMP data provide empirical validation of the inferred BLR emission to demonstrate the applicability of the method to real monitoring data.

\subsection{\texttt{STiSS}: SED Time Series Simulator}
\label{sec:mock_data}

We develop a SED time series Simulator (\texttt{STiSS}) to generate mock spectrophotometric data. For arbitrary choices of luminosity ($L_{5100}$), accretion rate ($\dot M$), and black hole mass ($M_{\rm BH}$), the \texttt{STiSS} combines emission from a central irradiating source, an accretion disk, and surrounding BLR clouds. Using the simulated SED for a mock AGN sample, we extract broad-band light curves through synthetic photometry. The details of the simulation are described below.

\subsubsection{Steady-state viscous thin disk}
\label{subsec:visc_disk}

We adopt a geometrically thin, optically thick Shakura--Sunyaev disk (SSD), for which the steady-state viscous temperature profile is given by \citep[following the expression of][]{Starkey2016}
\begin{equation}
\begin{split}
T_{\rm visc}^4(R)
=
\frac{3GM_{\rm BH}\dot{M}}{8\pi\sigma_{\rm SB}R^{3}}
\left(1-\sqrt{\frac{R_{\rm in}}{R}}\right)
\end{split}
\end{equation}
where $G$ is the gravitational constant, $\sigma_{\rm SB}$ is the Stefan--Boltzmann constant, and $R_{\rm in}$ is the inner disk radius, which is defined as the innermost stable circular orbit radius. The accretion rate is derived from the given luminosity of the AGN with a radiative efficiency of 0.1.

Disk-reverberation calculations often adopt a linear response formalism, in which the driving light curve is convolved with a wavelength-dependent response function \citep{Cackett2007},
\begin{equation}
\begin{aligned}
\psi(\tau, \lambda) = \iint &\frac{\partial B}{\partial T} \frac{\partial T}{\partial L} \frac{R dR d\theta \cos i}{D^2} \\
&\quad \times \delta\left(\tau - \frac{R}{c}(1+\sin i \cos\theta)\right)
\end{aligned}
\label{eq:disk-response}
\end{equation}
where $i$ is the viewing angle ($0^\circ$ for a face-on view), $\theta$ is the azimuthal angle around the rotational axis, $D$ is the distance between the observer and the disk, $L$ is the irradiating luminosity from the central source onto the disk, and $B$ is the blackbody radiation. Such a convolution-based treatment requires a linear approximation to the response of the Planck function and disk temperature to changes in the driving luminosity, through the factor $|(\partial B/\partial T)(\partial T/\partial L)|$ \citep[e.g.,][]{Cackett2007,Starkey2016,Kovacevic2022}. Instead of adopting this linearized response formalism, we explicitly compute the temperature fluctuations induced by the variable central irradiation and then synthesize the resulting time-dependent disk flux.

\subsubsection{Stochastic variability driver: damped random walk}
\label{subsec:drw_driver}

\begin{figure*}[ht!]
\centering
\includegraphics[width=\textwidth]{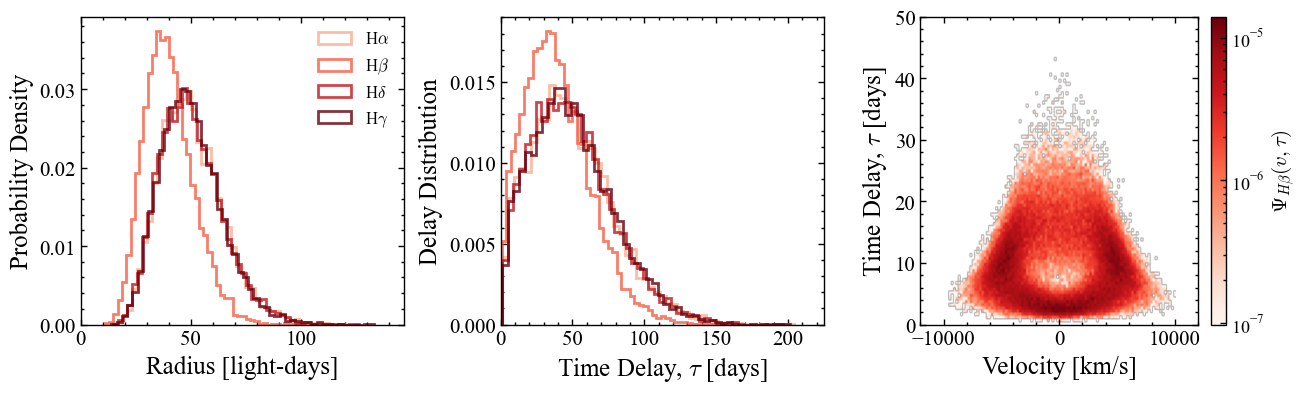}
\caption{Properties of the simulated BLR cloud ensemble used to generate time-dependent emission-line profiles. Left: the radial probability distributions of clouds associated with different emission lines. Center: the corresponding time-delay distributions after projecting the cloud geometry along the LOS. Right: the velocity--delay transfer function, $\Psi(v,\tau)$, which encodes the joint distribution of LOS velocity and reverberation delay for the modeled BLR response.}
\label{fig:blr_cloud_distributions}
\end{figure*}

The intrinsic variability is driven by a dimensionless stochastic process $X(t)$, modeled as an Ornstein--Uhlenbeck process, equivalent to a Damped Random Walk \citep[DRW;][]{Kelly2009,MacLeod2010} model. In a differential form, this process is written as
\begin{equation}
dX(t) = -\frac{X(t)}{\tau_{\rm DRW}}\,dt + \sigma_{\rm drive}\,dW(t)
\end{equation}
where $\tau_{\rm DRW}$ is the relaxation timescale, $dW(t)$ is the Wiener-process increment, and $\sigma_{\rm drive}$ is the stochastic driving amplitude.
\texttt{STiSS} uses $X(t)$ to specify the temporal structure of the stochastic fluctuations rather than their physical amplitude and each realization is subsequently normalized to zero mean and unit variance. The physical variability amplitude of the central irradiating source is instead introduced separately through $F_{\rm var}$ when mapping $X(t)$ to the irradiation luminosity.

Then, we express the time-dependent irradiation luminosity as
\begin{equation}
L_{\rm irr}(t)
=
\langle L_{\rm irr}\rangle\,
\exp\!\Big(F_{\rm var}\,X(t)\Big)
\end{equation}
where $\langle L_{\rm irr}\rangle$ sets the mean irradiation level and $F_{\rm var}$ controls the variability amplitude. Because the fractional variability is modified by thermal reprocessing and geometric smoothing in the accretion disk, $F_{\rm var}$ does not correspond directly to the optical variability amplitude. We therefore calibrate $F_{\rm var}$ through the full \texttt{STiSS} forward model so that the resulting optical-continuum variability matches the target $F_{\rm var}=0.2$ adopted for the mock data.

\subsubsection{Thermal reverberation from lamp-post irradiation}
\label{subsec:thermal_reverb}

We include thermal reprocessing of the variable irradiation by adopting a lamp-post geometry \citep{Cackett2007}, in which the driving source is located at height $h$ above the black hole. For a disk annulus at cylindrical radius $R$, the distance from the lamp to the disk is $r(R)=\sqrt{R^{2}+h^{2}}$. Thus, the incidence factor is approximated as $\cos\theta_{\rm inc}(R)=h/r(R)$.

We express the irradiation contribution to the disk temperature as
\begin{equation}
\begin{aligned}
T_{\rm irr}^{4}(R,t)
={}& \frac{(1-A)}{4\pi\sigma_{\rm SB}\,r(R)^{2}} \\
&\quad \times
L_{\rm irr}\!\Bigl(t-\tau(R)\Bigr)\,
\cos\theta_{\rm inc}(R)
\end{aligned}
\end{equation}
where $A$ is the disk albedo and the source-to-disk light-travel delay is $\tau(R)=r(R)/c$. The total temperature profile,
\begin{equation}
T(R,t)=\Big[T_{\rm visc}^{4}(R)+T_{\rm irr}^{4}(R,t)\Big]^{1/4}
\end{equation}
fluctuates over time as shown in Figure~\ref{fig:disk_temperature_snapshots}. The time-dependent spectral flux density is obtained by integrating the blackbody emission over the time-dependent temperature profile as,
\begin{equation}
\begin{split}
f_{\lambda}(\lambda,t)
=
\frac{\cos i_{\rm disk}}{D^{2}}
\int_{R_{\rm in}}^{R_{\rm out}} B_{\lambda}\!\Big(T(R,t)\Big)\, 2\pi R\,dR
\end{split}
\end{equation}
where $i_{\rm disk}$ is the disk inclination, $D$ is the luminosity distance, $R_{\rm out}$ is the outer disk radius, and $B_{\lambda}(T)$ is the Planck function. This integration yields a time-variable power-law continuum.

\subsubsection{BLR model}
\label{subsec:blr}

We model the BLR as an ensemble of discrete, non-interacting clouds, adopting a simplified version of the \texttt{CARAMEL} framework \citep{Pancoast2014}, in which inflows and outflows are neglected and the clouds trace a static, virialized distribution.

Cloud radii are drawn from a shifted-gamma distribution,
\begin{equation}
R_{\rm cloud} \sim {\rm \Gamma}(\beta,\theta) + R_{\rm shift}
\end{equation}
where $\beta$ is the shape parameter, $\theta$ the scale parameter, and $R_{\rm shift}$ a hard inner offset. The scale $\theta$ is fixed by requiring the ensemble mean radius to equal the reverberation radius $R_{\rm BLR}$ from the size--luminosity ($RL$) relation, $R_{\rm BLR}\propto L_{5100}^{0.5}$ \citep{Bentz2013,Woo2024}, evaluated at the mean $5100\,\text{\AA}$ continuum luminosity.

\begin{table}[t]
\centering
\caption{Fixed parameters adopted in \texttt{STiSS}.}
\small
\begin{tabular*}{\columnwidth}{@{\extracolsep{\fill}}lll@{}}\hline\hline
Parameter & Symbol & Value \\
\hline
\multicolumn{3}{@{}l}{\textit{Driver}} \\
Lamp-post height & $h$ & $40\,R_{\rm g}$ \\
Driver variability amplitude & $F_{\rm var}$ & $0.2$ \\
DRW relaxation timescale & $\tau_{\rm DRW}$ & $150~{\rm days}$ \\
\hline
\multicolumn{3}{@{}l}{\textit{Disk}} \\
Albedo & $A$ & $0.3$ \\
Inclination & $i_{\rm disk}$ & $45^\circ$ \\
Outer radius & $R_{\rm out}$ & $3000\,R_{\rm g}$ \\
Mean irradiation luminosity & $\langle L_{\rm irr} \rangle$ & $0.85\,L_{\rm bol}$ \\
\hline
\multicolumn{3}{@{}l}{\textit{BLR}} \\
Opening angle & $\theta_{\rm o}$ & $45^\circ$ \\
Inclination & $i_{\rm BLR}$ & $45^\circ$ \\
Shape parameter & $\beta$ & $10$ \\
Macroturbulence & $\sigma_{\rm turb}$ & $0.3$ \\
Radial scaling factors & $(s_{\rm H\alpha}, s_{\rm Fe})$ & $(1.25, 1.5)$ \\
Responsivity index & $\gamma$ & $0.5$ \\
\hline
\end{tabular*}
{\raggedright\tablecomments{The radial scaling factor is defined relative to H$\beta$, such that $s_X = 1$ for H$\beta$.}}
\label{tab:sim_param}
\end{table}

\begin{figure*}[ht!]
\centering
\includegraphics[width=\textwidth]{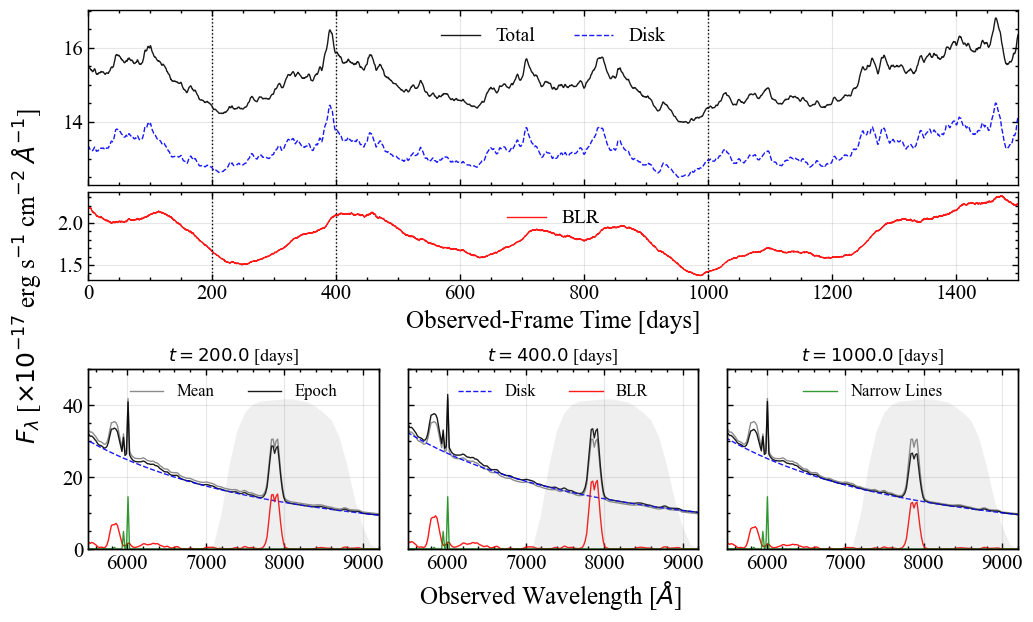}
\caption{An illustrative demonstration of \texttt{STiSS}. Top: Total (black), disk (blue), and BLR (red) light curves obtained by integrating the time-dependent spectra through the $I$-band filter. The vertical dotted lines mark the epochs shown in the bottom panels. Bottom: snapshots of the synthetic spectra at three representative epochs, compared with the time-averaged spectrum (gray). The disk (blue), BLR (red), and narrow-line (green) components are shown separately, and the shaded region indicates the $I$-band filter transmission curve used for the synthetic photometry.}
\label{fig:mock_spectral_timeseries}
\end{figure*}

The clouds populate a thick disk of half-opening angle $\theta_{\rm o}$. Measuring the polar angle ($\vartheta$) from the rotational axis, the clouds are distributed uniformly in $\cos\vartheta$ within $\theta_{\rm o}$,
\begin{equation}
\cos\vartheta \sim \mathcal{U}\!\left[-\sin\theta_{\rm o},\,+\sin\theta_{\rm o}\right],
\qquad \phi \sim \mathcal{U}\!\left[0,2\pi\right]
\end{equation}
so that $\theta_{\rm o}\!\to\!0^{\circ}$ recovers a thin equatorial ring and $\theta_{\rm o}\!\to\!90^{\circ}$ a spherical distribution. The cloud positions are then rotated by the BLR inclination ($i_{\rm BLR}$) relative to the line of sight (LOS). Emission is weighted toward the near side of the flow by a linear anisotropy factor $w=\tfrac{1}{2}\left[\,1+\zeta\,(z/R_{\rm cloud})\,\right]$, where $z$ is the LOS coordinate and $\zeta\in[0,1]$ sets the near/far asymmetry.

Each cloud is assigned a LOS velocity composed of Keplerian rotation plus macroturbulence,
\begin{align}
v_{\rm los} &= v_{\rm circ}(r)\,\sin i_{\rm BLR}\,\cos\phi + v_{\rm turb} \\
v_{\rm turb} &\sim \mathcal{N}\!\Big(0,\ \sigma_{\rm turb}\,v_{\rm circ}(r)\Big)
\end{align}
with $v_{\rm circ}(r)=\sqrt{GM_{\rm BH}/r}$ and $\sigma_{\rm turb}$ the relative macroturbulent amplitude.

Rather than share a single response across all transitions, each line $X$ is assigned its own cloud ensemble whose mean radius is scaled from the H$\beta$ radius by a fixed ratio,
\begin{equation}
R_{X} = s_{X}\,R_{\rm BLR}(\mathrm{H}\beta)
\label{eq:radial_scaling}
\end{equation}
where the $s_{X}$ encode the radial stratification of the BLR. For each line, the ensemble is re-sampled at radius $R_{X}$ while sharing the same dimensionless geometry ($\beta,\theta_{\rm o},i,\zeta$) and kinematics. The geometric delay of each cloud, $\tau = (r-\vec{r}\cdot\hat{n}_{\rm obs})/c$, together with $v_{\rm los}$, defines the line-specific velocity--delay response $\Psi_{X}(v,\tau)$ (example shown in Figure~\ref{fig:blr_cloud_distributions}); the line light curve is obtained by convolving the responsivity-weighted continuum with $\Psi_{X}$, and the line profile is the sum of Doppler-shifted cloud contributions at delayed time $t-\tau$. Because the ensembles differ only in their radial scale, the resulting transfer functions form a virially self-similar family related by $\tau\!\to\!s_{X}\tau$ and $v\!\to\!v/\sqrt{s_{X}}$, i.e., each line has a distinct delay distribution set by its size, while sharing a common normalized velocity--delay structure.

The transfer function, together with the assumption that the line emission responds non-linearly to the continuum as $L_{\rm line}(t)\propto L_{\rm cont}(t)^{\gamma}$ with responsivity index $\gamma$, yields time-dependent emission line profiles.

\begin{figure*}[!t]
\centering
\includegraphics[width=\textwidth]{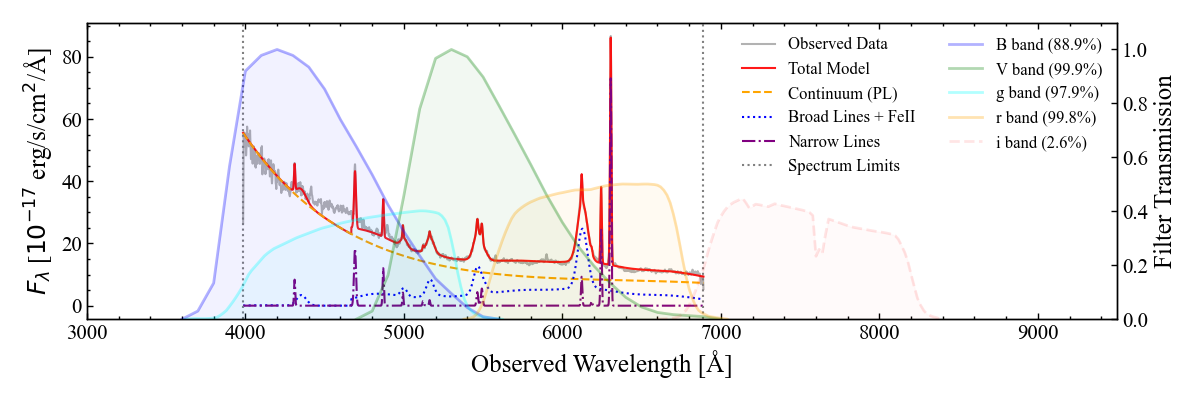}
\caption{
Example single-epoch spectrum of an SAMP AGN J1026+523 (gray) and the best-fit model (red). The spectrum is decomposed into a power-law continuum (orange), broad-line plus Fe~\textsc{ii} emission (blue), and narrow-line (purple) components. The overlaid curves show the transmission functions of the broad-band filters used for the photometric monitoring. The fractional overlap between each filter transmission curve and the spectrum is indicated in the legend.
}
\label{fig:samp_spectral_decomposition}
\end{figure*}

\subsubsection{Additional spectral components}
\label{subsec:additional_components}

\paragraph{Fe\,\textsc{ii} pseudo-continuum.}
We model the optical Fe\,\textsc{ii} pseudo-continuum using the empirical template of \citet{Boroson1992}. The Fe\,\textsc{ii} response radius is determined by applying a radial scaling factor in the same manner as Eq.~\ref{eq:radial_scaling}. Assuming virial motion at this radius, we compute the corresponding velocity width from the black hole mass and use this value as the Gaussian broadening width $\sigma_{\rm Fe}$ when convolving the template before including it in the synthetic spectrum.

\paragraph{Narrow emission lines.}
Narrow lines (e.g., [O\,\textsc{iii}]) are added as Gaussian profiles with fixed widths of 500km/s. They are treated as constant on the simulation timescales and serve as flux-calibration references.

\subsubsection{Synthetic broad-band light curves}
\label{subsec:sim_grid_and_synth}

\texttt{STiSS} can be widely used for various purposes, i.e., simulation-based inference. In this work, we utilize \texttt{STiSS} to generate synthetic datasets for mock AGN spanning $\log \left( L_{5100} / {\rm erg~s^{-1}} \right) = \{41, 42, 43, 44, 45, 46\}$ and $\lambda_{\rm Edd} = \{0.01, 0.1, 1.0\}$, while the other parameters are fixed (see Table~\ref{tab:sim_param}). For each AGN, we produce time-dependent spectra following the procedures described above and extract synthetic broad-band light curves after shifting the spectra to the observed frame at $z=0.2$ and using the $BVRI$ filter transmission curves (see an example in Figure~\ref{fig:mock_spectral_timeseries}). We choose $z=0.2$ to place the H$\beta$ and H$\alpha$ lines in the V and I bands, respectively.

\subsection{SAMP}
\label{subsec:samp_data_details}

We use the light curves from SAMP, which monitored moderate- to high-luminosity AGN over 2015--2021, with typical cadences of $\sim$3--5 days for photometry and $\sim$20--30 days for spectroscopy \citep{Woo2019}. The H$\beta$ lags were reported for 32 AGN by \citet{Woo2024}. However, we use 19 targets for which the required multi-band photometric light curves, optical spectra, and spectroscopic H$\beta$ light curves are available. The broad-band light curves are available in $B$, $g$, $V$, $r$, and $i$ \citep{Mandal2025} bands with effective wavelengths of 4450\AA, 4770\AA, 5510\AA, 6231\AA, and 6231\AA, respectively. These light curves cover broad wavelength coverage needed to constrain the wavelength-dependent disk response, while the spectroscopic H$\beta$ light curves can be used to testing the inferred BLR emission by \texttt{TILDE}. We present an example of a SAMP AGN, J1026+523, in Figure~\ref{fig:samp_spectral_decomposition}, illustrating the significant contribution of broad lines, Fe~{\sc ii}, and narrow-line components. 

\begin{figure*}[ht!]
\centering
\includegraphics[width=\textwidth]{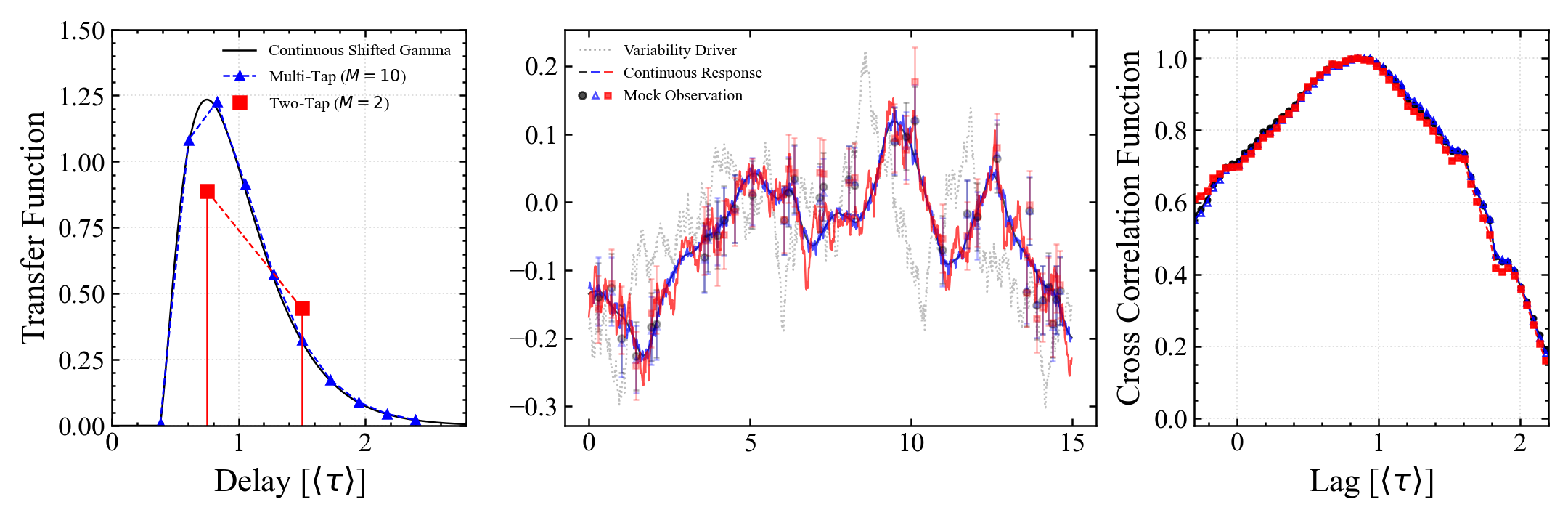}
\caption{Comparison of a continuous shifted-gamma transfer function (black) with its multi-tap ($M=10$; blue) and two-tap ($M=2$; red) approximations with all time axes in units of transfer function centroid. Left: the three transfer functions where centroids are aligned. Center: response light curves produced by convolving each transfer function with an example DRW variability driver (gray). Solid curves show the continuous responses, while markers show mock observations generated through irregular sampling and the addition of effective noise, assuming a broad-band measurement uncertainty of 1.5\% and $\xi=0.2$. Right: cross-correlation functions between the variability driver and the three mock response light curves.}
\label{fig:two-tap_comparison}
\end{figure*}


\section{\texttt{TILDE}: Two-stage Inference for Light curve DEcomposition}
\label{sec:method}

\texttt{TILDE} is a Bayesian framework that separates AGN broad-band photometric variability into three additive contributions: a static component from the host galaxy and narrow-line emission, a variable accretion-disk continuum, and a variable BLR. With such a forward model, the decomposition is performed in two stages. In Stage~1, we infer the structural and band-local parameters of a joint forward model under a multivariate Gaussian photometric likelihood, regularized by spectroscopic priors obtained from a single-epoch optical spectrum. In Stage~2, we condition a Gaussian-process (GP) posterior on the Stage~1 samples and analytically reconstruct the latent disk-continuum and BLR-contamination light curves over a dense epoch grid.

\subsection{Photometric Forward Model}
\label{sec:forward}

\texttt{STiSS}, described in \S\ref{sec:mock_data}, provides a physically motivated representation of the disk and BLR variability, but we do not use this full simulator directly as the forward model for fitting the observed broad-band light curves. In broad-band photometry, many of the parameters project onto nearly indistinguishable observables and the resulting posterior would therefore be strongly degenerate, computationally expensive to explore, and sensitive to assumptions about microscopic disk and BLR geometry that are not uniquely constrained by the data. We instead use \texttt{STiSS} as a controlled source of mock data with known disk and BLR components, and develop a reduced but statistically stable forward model for inference. 

For filter band $b$ at observed time $t$, we model the measured flux as
\begin{equation}
\begin{aligned}
F_b(t) ={}& \bar F_{\mathrm{cont},b}\,\bigl[1+c_b(t)\bigr] \\
&\quad
+ \xi_b\,\bar F_{\mathrm{cont},b}\, \bigl[1+\alpha\,l_b(t)\bigr] + F_{\mathrm{static},b}
\end{aligned}
\label{eq:forward}
\end{equation}
where $F_{\mathrm{static},b}$ is the time-independent host+narrow-line component, $\bar F_{\mathrm{cont},b}$ is the band-averaged disk continuum flux, $\xi_b$ is the BLR-to-continuum flux ratio in band $b$, $\alpha$ is a global scaling between continuum and BLR fractional variability amplitudes, and $c_b(t)$ and $l_b(t)$ are zero-mean fractional fluctuations of the disk continuum and BLR-contamination responses, respectively. Both $c_b$ and $l_b$ derive from a single latent driving process $X(t)$ through the band-specific response operators defined below.

We model $X(t)$ as a DRW with stationary covariance kernel
\begin{equation}
K_{\mathrm{DRW}}(\Delta t) = \sigma_{\mathrm{DRW}}^{2}
\exp\left(-\frac{|\Delta t|}{\tau_{\mathrm{DRW}}}\right)
\label{eq:drw}
\end{equation}
with $\sigma_{\mathrm{DRW}}$ and $\tau_{\mathrm{DRW}}$ sampled in log space.

\paragraph{Disk response}
In a continuous response formulation, the fractional disk-continuum fluctuation in band $b$ is obtained by convolving the latent driving process with a band-dependent disk transfer function,
\begin{equation}
c_b(t) = \int_{0}^{\infty} \Psi_{\mathrm{cont},b}(\delta t)\, X(t-\delta t)\,d\delta t
\label{eq:disk-conv}
\end{equation}
where $\Psi_{\mathrm{cont},b}(\delta t)$ describes the distribution of light-travel delays contributing to the observed continuum flux in that band. We take the transfer function to be normalized, $\int_{0}^{\infty}\Psi_{\mathrm{cont},b}(\delta t)\,d\delta t=1$, so that $c_b(t)$ remains a zero-mean fractional response when the driver $X(t)$ is zero mean.

Rather than evaluating a full disk-integrated radiative-transfer response, we approximate Eq.~\eqref{eq:disk-conv} with a $M=2$ discrete kernel, i.e., two-tap approximation (see an example in Figure ~\ref{fig:two-tap_comparison}), 
\begin{equation}
\begin{aligned}
\Psi_{\mathrm{cont},b}(\delta t)
\simeq{}&
(1-\eta_c)\,\delta_{\mathrm{D}}\!\left(\delta t-\tau^{(s)}_{\mathrm{cont},b}\right) \\
&\quad
+
\eta_c\,\delta_{\mathrm{D}}\!\left[\delta t-(1+f)\tau^{(s)}_{\mathrm{cont},b}\right]
\end{aligned}
\label{eq:disk-k2-kernel}
\end{equation}
where $\eta_c$ and $f_c$ are two-tap weights for disk and $\delta_{\mathrm{D}}$ is the delta function. The superscript $(s)$ denotes a sampled delay parameter, distinguished from the centroid lags defined below. Substituting Eq.~\eqref{eq:disk-k2-kernel} into Eq.~\eqref{eq:disk-conv} gives
\begin{equation}
\begin{aligned}
c_b(t)
={}& (1-\eta_c)\,X\!\bigl(t-\tau^{(s)}_{\mathrm{cont},b}\bigr) \\
&\quad
+ \eta_c\,X\!\bigl[t-(1+f_c)\tau^{(s)}_{\mathrm{cont},b}\bigr]
\end{aligned}
\label{eq:disk-k2}
\end{equation}

\begin{table}
\centering
\caption{Priors for the \texttt{TILDE} parameters}
\small
\begin{tabular*}{\columnwidth}{@{\extracolsep{\fill}}lll@{}}
\hline\hline
Parameter & Prior & Physical Bounds \\
\hline
$\log\alpha$                 & LogUniform                                                                 & $(0,\,5]$ \\
$\log\sigma_{\rm DRW}$       & LogUniform                                                                 & $>0$ \\
$\log\tau_{\rm DRW}$         & LogUniform                                                                 & $>0$ \\
$\log A_{\rm disk}$          & LogNormal$(\ln(1),\,2^2)$                                                  & $[0.01,\,300]$ \\
$\beta_{\rm disk}$           & Normal$(4/3,\,0.2^2)$                                                      & $[0.1,\,3]$ \\
$\log s_{\rm phot}$          & Normal$(0.01,\,0.01^2)$                                                    & $[0.001,\,0.05]$ \\
\hline
$\log\bar{F}_{{\rm cont},b}$ & Normal$(F_{\mathrm{cont},b}^{\rm spec}, \sigma^2_{F\mathrm{cont}, b})$     & $>0$ \\
$\log\xi_{b}$                & Normal$(\xi_{b}^{\rm spec}, \sigma^2_{\xi, b})$                            & $>0$ \\
$F_{{\rm static},b}$         & Normal$(F_{\mathrm{static},b}^{\rm spec}, \sigma^2_{\mathrm{static}, b})$ & $>0$ \\
$\log\tau_{{\rm line},b}$    & LogNormal$(\tau_{\rm line}^{\rm spec}, \sigma^2_{RL})$                     & $(0,\,1000]$ \\
$\log R_{F_{\rm var}}$       & LogNormal$(\ln(0.5),\, 0.5^2)$                                             & $>0$ \\
\hline
\end{tabular*}
{\raggedright\tablecomments{(Left column): parameter names of \texttt{TILDE}, where ``$\log X$'' indicates that the parameter is sampled in natural log-space ($\ln X$). Prior distributions (middle column) and hard boundaries (right column) are quoted in terms of the physical parameter $X$. Intervals with open parentheses (e.g., $(0, 5]$) denote strictly positive physical parameters where $X > 0$ ($\ln X > -\infty$).}}
\label{tab:tilde_priors}
\end{table}

We hold $\eta_c=0.5$ and $f_c=1$ fixed throughout this work. This $M=2$ approximation is a tractable quadrature of the disk response: it allows the continuum light curve to be broadened relative to a pure delta-function lag while preserving an analytic covariance under the DRW driver. It should therefore be interpreted as an effective two-point approximation to the disk transfer function, not as a radiative-transfer integration over disk annuli. We do not adopt higher-order approximations because they increase computational cost without providing additional useful information, as differences between the $M=2$ and $M\geq3$ cases are not resolvable within the observational uncertainties (see Figure ~\ref{fig:two-tap_comparison}). 

With the $M=2$ approximation, the physically meaningful centroid lag is $\tau_{\rm cont,b} = 1.5\,\tau^{(s)}_{\rm cont,b}$. This constant factor follows from the centroid of the adopted symmetric two-tap continuum kernel, which assigns equal weights to delays of $\tau^{(s)}_{\rm cont,b}$ and $2\tau^{(s)}_{\rm cont,b}$. The band-specific continuum lag scale follows the standard thin-disk wavelength dependence,
\begin{equation}
\tau_{\mathrm{cont},b} = 1.5\tau^{(s)}_{\mathrm{cont},b} \;=\; A_{\mathrm{disk}}\,\left[\left(\frac{\lambda_b}{\lambda_{\mathrm{ref}}}\right)^{\beta_{\mathrm{disk}}}-1\right]
\label{eq:taucont}
\end{equation}
with $A_{\mathrm{disk}}$ and $\beta_{\mathrm{disk}}$ sampled. 

\paragraph{BLR response}
We model the BLR response with the same two-tap discrete delay kernel used for the disk,
\begin{equation}
\begin{aligned}
l_b(t)
={}& (1-\eta_{\ell})\,X\!\bigl(t-\tau^{(s)}_{\mathrm{line},b}\bigr) \\
&\quad
+ \eta_{\ell}\,X\!\bigl[t-(1 + f_{\ell})\,\tau^{(s)}_{\mathrm{line},b}\bigr]
\end{aligned}
\label{eq:blr-k2}
\end{equation}
with $\eta_{\ell}=1/3$ and $f_{\ell}=1$ held fixed, and $\tau^{(s)}_{\mathrm{line},b}$ inferred independently per band. The two-tap weighting is asymmetric: it places two thirds of the response at the primary delay $\tau^{(s)}_{\mathrm{line},b}$ and one third at $2\,\tau^{(s)}_{\mathrm{line},b}$, producing a forward-skewed effective transfer function whose first-moment lag equals $\tau_{\mathrm{line},b} = (4/3)\,\tau^{(s)}_{\mathrm{line},b}$. The sampled parameter is therefore the position of the leading delta, smaller by a factor of $3/4$ than the broad-band CCF centroid one would measure from the same light curves; this multiplicative offset must be accounted for when comparing to reported reverberation measurements. The choice $\eta_{\ell}=1/3$ rather than the symmetric $\eta_c=1/2$ used for the disk reflects the typical \emph{shape} of broad-line transfer functions, which are skewed toward shorter lags due to the geometric weighting of near-side BLR clouds; we adopt these values as fixed hyperparameters and do not infer them. Because a broad-band filter intercepts a band-specific mixture of broad emission lines, \ion{Fe}{2} pseudo-continuum, and diffuse continuum radiation, a single global line lag cannot reproduce the per-band reverberation signature; we interpret $\tau_{\mathrm{line},b}$ as an \emph{effective broad-band BLR lag}, not as the lag of any individual emission line.

\subsection{Spectroscopic Constraints}
\label{sec:spec}

For each target, we use a single epoch spectrum to anchor the multiplicative flux decomposition in Stage~1. The spectrum, $f(\lambda)$, is decomposed into three components with \texttt{PyQSOFit} \citep{Guo2018,Shen2019}: a power-law continuum component ($f_{\rm cont}$); a line component ($f_{\rm line}$), including \ion{Fe}{2} emission, Balmer continuum, and broad emission lines; and a static component ($f_{\rm static}$), including host-galaxy emission and narrow emission lines, as shown in Figure~\ref{fig:samp_spectral_decomposition}. Each component is then synthesized per band, $b$, through the filter transmission curve, $T_b(\lambda)$,
\begin{equation}
F_{\mathrm{(comp)}, b}^{\mathrm{spec}} =
\frac{\int T_b(\lambda)\lambda f_{(\mathrm{comp})}(\lambda)d\lambda}
{\int T_b(\lambda)\lambda d\lambda}
\label{eq:bandflux}
\end{equation}
yielding the continuum level ($F_{\mathrm{cont},b}^{\mathrm{spec}}$), line-to-continuum ratio ($\xi_b^{\mathrm{spec}} \equiv F_{\mathrm{line},b}^{\rm spec}/F_{\mathrm{cont},b}^{\rm spec}$), and the static (non-variable) flux ($F_{\mathrm{static},b}^{\mathrm{spec}}$). We also use the rest-frame monochromatic continuum luminosity ($\lambda L_{5100}$), measured from the fitted AGN continuum, to estimate the characteristic ($\mathrm H\beta$) BLR radius from the empirical $RL$ relation of \citet{Woo2024},
\begin{equation}
\log \left[\frac{\tau_{\mathrm{line}}^{\rm spec}}{(\rm days)}\right] = 1.41 + 0.40 \log \left[\frac{\lambda L_{5100}}{(10^{44} \mathrm{erg} \mathrm{s}^{-1})}\right]
\end{equation}
with intrinsic scatter $\sigma_{RL}=0.177$. These measurements, together with uncertainties estimated from MC perturbations and linear error propagation where required, enter Stage~1 as spectroscopic priors or regularization terms (\S\ref{sec:stage1-posterior}), rather than fixed constraints. We assess whether the individual SAMP effective BLR-lag posteriors remain tied to, or depart from, this adopted $RL$ regularization through a prior-to-posterior comparison in Appendix~\ref{app:prior_posterior_lags}. We treat both $\xi_b$ and the host contribution as static over the duration of the light curve.

\subsection{Stage 1: Bayesian Inference of Structural Parameters}
\label{sec:stage1}

Stage~1 infers the joint posterior over the forward model (described in \S\ref{sec:forward}) parameter vector that we split into a global block and a band-local block.

\subsubsection{Parameter vector}
\label{sec:stage1-pars}

The six global parameters control the latent driver and the disk/BLR response geometry,
\begin{equation}
\boldsymbol{\theta}_{\mathrm{glob}} \;=\;
   \bigl\{\,\alpha,\;\sigma_{\mathrm{DRW}},\;\tau_{\mathrm{DRW}},\;A_{\mathrm{disk}},\;\beta_{\mathrm{disk}},\;s_{\mathrm{phot}}\bigr\}
\label{eq:theta-glob}
\end{equation}
where $s_{\mathrm{phot}}$ is an intrinsic fractional photometric jitter (\S\ref{sec:stage1-like}). For each band $b \in \mathcal{B}$, we carry four band-local parameters,
\begin{equation}
\boldsymbol{\theta}_b \;=\;
   \bigl\{\,\bar F_{\mathrm{cont},b},\;\xi_b,\;F_{\mathrm{static},b},\;\tau^{(s)}_{\mathrm{line},b}\bigr\}
\label{eq:theta-band}
\end{equation}
so that the full sampled vector is $\boldsymbol{\theta}=\boldsymbol{\theta}_{\mathrm{glob}} \cup
\bigcup_{b\in\mathcal{B}}\boldsymbol{\theta}_b$, of dimension $6+4|\mathcal{B}|$. The band-specific continuum lag $\tau_{\mathrm{cont},b}$ is not sampled; it is deterministically derived from $(A_{\mathrm{disk}},\beta_{\mathrm{disk}})$ via Eq.~\eqref{eq:taucont}.

\subsubsection{Photometric likelihood}
\label{sec:stage1-like}

Let $\mathbf{Y}$ be the concatenated flux vector stacked band-by-band across all observed epochs, $\mathbf{Y}_{\mathrm{err}}$ the corresponding measurement errors, and $\boldsymbol{\mu}(\boldsymbol{\theta})$ the model mean implied by Eq.~\eqref{eq:forward}. The joint covariance matrix $\mathbf{C}(\boldsymbol{\theta})$ is assembled as a sum of three contributions,
\begin{equation}
\begin{aligned}
\mathbf{C}(\boldsymbol{\theta})
={}& \mathbf{C}_{\mathrm{latent}}(\boldsymbol{\theta}) \\
&\quad + \operatorname{diag}(\mathbf{Y}_{\mathrm{err}}^{\,2})+ \operatorname{diag}\!\left[(s_{\mathrm{phot}}\,\bar F_b)^{2}\right]
\end{aligned}
\label{eq:Ctotal}
\end{equation}
where $\mathbf{C}_{\mathrm{latent}}$ is obtained by propagating the DRW kernel (Eq.~\ref{eq:drw}) through the two-tap continuum (Eq.~\ref{eq:disk-k2}) and BLR (Eq.~\ref{eq:blr-k2}) responses. For compactness, we define the latent covariance entry and the DRW covariance kernel as
\begin{align}
C_{ij}^{bb'}
&\equiv
\left[\mathbf{C}_{\mathrm{latent}}\right]_{(b,i),(b',j)} \label{eq:gp-mean}\\
\mathcal{K}_{x}^{pq} 
& \equiv
K_{\mathrm{DRW}}\!\left(
\Delta t_{ij}-\Delta\tau_{x}^{pq}
\right)   
\label{eq:compact}
\end{align}
where $x\in\{c,\ell,c\ell,\ell c\}$. Then, for a pair of observations $(b,t_i)$ and $(b',t_j)$, the corresponding latent covariance entry is
\begin{equation}
\begin{aligned}
C_{ij}^{bb'}
={}&
A_b A_{b'}
\sum_{p,q \in \{0,1\}}
w_{c,p} w_{c,q}\,
\mathcal{K}_{c}^{pq} \\
&+
B_b B_{b'}
\sum_{p,q \in \{0,1\}}
w_{\ell,p} w_{\ell,q}\,
\mathcal{K}_{\ell}^{pq} \\
&+
A_b B_{b'}
\sum_{p,q \in \{0,1\}}
w_{c,p} w_{\ell,q}\,
\mathcal{K}_{c\ell}^{pq} \\
&+
B_b A_{b'}
\sum_{p,q \in \{0,1\}}
w_{\ell,p} w_{c,q}\,
\mathcal{K}_{\ell c}^{pq}.
\end{aligned}
\label{eq:Clatent}
\end{equation}
Here $A_b\equiv \bar F_{\mathrm{cont},b}$ and
$B_b\equiv \bar F_{\mathrm{cont},b}\xi_b\alpha$. The $M=2$ tap weights are
defined as
\begin{equation}
w_{c,p}=(1-\eta_c)^{1-p}\eta_c^p,
\qquad
w_{\ell,p}=(1-\eta_\ell)^{1-p}\eta_\ell^p
\end{equation}
with $\eta_c=1/2$ and $\eta_\ell=1/3$. The quantities $\Delta\tau_{c}^{pq}$, $\Delta\tau_{\ell}^{pq}$, $\Delta\tau_{c\ell}^{pq}$, and $\Delta\tau_{\ell c}^{pq}$ denote the corresponding differences between the continuum and BLR lags evaluated at the two $M=2$ taps for the two bands. The measurement-noise and photometric-jitter diagonal terms in Eq.~\eqref{eq:Ctotal} use the per-band mean flux $\bar F_b$, so that $s_{\mathrm{phot}}$ represents a fractional calibration floor.

The photometric log-likelihood, $\ln\mathcal L_{\mathrm{phot}}$, is the standard multivariate Gaussian
\begin{equation}
\begin{aligned}
\ln\mathcal{L}_{\mathrm{phot}}(\boldsymbol{\theta})
={}& -\tfrac{1}{2}\Bigl[N\ln(2\pi) + \ln\det\mathbf{C} \\
&\quad + (\mathbf{Y}-\boldsymbol{\mu})^{\mathsf{T}}\mathbf{C}^{-1}(\mathbf{Y}-\boldsymbol{\mu})\Bigr]
\end{aligned}
\label{eq:likelihood}
\end{equation}
evaluated by Cholesky factorization of $\mathbf{C}$ with a small diagonal jitter added as a fallback on ill-conditioned matrices.

\subsubsection{Posterior and spectroscopic regularization}
\label{sec:stage1-posterior}

\begin{table*}[t]
\centering
\caption{Results of \texttt{TILDE} Inferred Disk and BLR Sizes}
\small
\begin{tabular*}{\textwidth}{@{\extracolsep{\fill}}lccccc@{}}
\hline\hline
\multicolumn{1}{c}{Object}
& $R_{5100,\mathrm{SSD}}$
& $R_{5100,\mathrm{ICCF}}$
& $R_{5100,\mathrm{TILDE}}$
& $\tau_{\mathrm{RM},\mathrm{H}\beta}$
& $\tau_{\mathrm{TILDE},\mathrm{line}}$ \\
& (light-days)
& (light-days)
& (light-days)
& (days)
& (days) \\
\multicolumn{1}{c}{(1)}
& (2)
& (3)
& (4)
& (5)
& (6) \\
\hline
Mrk 1501 & $0.83^{+0.15}_{-0.34}$ & $8.02 \pm 0.26$ & $0.43^{+0.61}_{-0.34}$ & $11.70^{+7.60}_{-8.90}$ & $33.70^{+11.65}_{-7.07}$ \\
PG 0026+129 & $1.42^{+0.78}_{-0.43}$ & $22.25 \pm 7.50$ & $22.51^{+4.11}_{-3.83}$ & $13.00^{+30.20}_{-18.50}$ & $88.79^{+26.90}_{-24.16}$ \\
PG 0052+251 & $2.35^{+0.16}_{-0.18}$ & $18.63 \pm 2.07$ & $3.47^{+3.24}_{-2.10}$ & $63.90^{+12.00}_{-10.90}$ & $67.12^{+19.46}_{-14.46}$ \\
J0101+422 & $2.83^{+0.18}_{-0.20}$ & $13.71 \pm 1.29$ & $3.15^{+1.87}_{-1.86}$ & $76.10^{+13.20}_{-11.80}$ & $59.75^{+17.33}_{-12.99}$ \\
PG 0947+396 & $1.84^{+0.17}_{-0.25}$ & $10.61 \pm 0.52$ & $0.53^{+0.94}_{-0.41}$ & $36.70^{+9.50}_{-11.00}$ & $62.61^{+17.57}_{-12.69}$ \\
J1026+523 & $1.06^{+0.08}_{-0.10}$ & $9.06 \pm 2.59$ & $0.85^{+0.86}_{-0.66}$ & $34.10^{+4.20}_{-4.00}$ & $37.22^{+3.14}_{-4.90}$ \\
J1120+423 & $1.81^{+0.22}_{-0.27}$ & $9.06 \pm 1.03$ & $0.64^{+1.06}_{-0.49}$ & $44.40^{+15.60}_{-14.90}$ & $47.03^{+10.35}_{-10.62}$ \\
PG 1121+422 & $1.95^{+0.16}_{-0.13}$ & $12.42 \pm 1.55$ & $1.77^{+4.25}_{-1.52}$ & $115.80^{+24.20}_{-20.20}$ & $70.73^{+27.25}_{-18.20}$ \\
PG 1202+281 & $1.48^{+0.12}_{-0.13}$ & $7.24 \pm 2.59$ & $0.94^{+1.57}_{-0.74}$ & $38.50^{+9.10}_{-8.50}$ & $55.32^{+13.60}_{-13.38}$ \\
J1217+333 & $1.01^{+0.21}_{-0.34}$ & $6.99 \pm 4.66$ & $0.55^{+0.84}_{-0.45}$ & $26.50^{+21.20}_{-20.70}$ & $35.70^{+13.89}_{-8.28}$ \\
VIII Zw 218 & $1.65^{+0.13}_{-0.15}$ & $18.37 \pm 2.59$ & $0.47^{+1.01}_{-0.36}$ & $63.30^{+16.40}_{-15.40}$ & $39.30^{+13.31}_{-8.72}$ \\
PG 1322+659 & $1.91^{+0.21}_{-0.26}$ & $8.80 \pm 0.52$ & $3.76^{+6.85}_{-3.28}$ & $49.30^{+19.20}_{-16.60}$ & $29.99^{+7.31}_{-5.51}$ \\
J1415+483 & $1.51^{+0.22}_{-0.22}$ & $18.11 \pm 2.33$ & $1.28^{+9.38}_{-1.12}$ & $25.30^{+11.80}_{-11.10}$ & $53.78^{+12.13}_{-10.73}$ \\
PG 1440+356 & $1.12^{+0.17}_{-0.23}$ & $20.96 \pm 8.28$ & $3.18^{+11.72}_{-2.80}$ & $50.60^{+16.90}_{-20.90}$ & $13.86^{+3.51}_{-2.76}$ \\
J1456+380 & $2.81^{+0.15}_{-0.16}$ & $14.23 \pm 4.40$ & $0.51^{+0.72}_{-0.38}$ & $77.80^{+9.40}_{-8.70}$ & $65.15^{+13.20}_{-18.14}$ \\
J1526+275 & $3.00^{+0.27}_{-0.51}$ & $16.30 \pm 1.81$ & $0.77^{+2.14}_{-0.63}$ & $63.90^{+10.30}_{-9.30}$ & $43.14^{+6.87}_{-4.45}$ \\
J1540+355 & $1.18^{+0.12}_{-0.15}$ & $5.43 \pm 3.88$ & $5.61^{+1.73}_{-1.29}$ & $57.90^{+18.10}_{-14.60}$ & $43.98^{+12.67}_{-10.44}$ \\
J1619+501 & $1.66^{+0.12}_{-0.13}$ & $4.14 \pm 5.95$ & $2.03^{+1.26}_{-1.41}$ & $32.30^{+6.80}_{-6.70}$ & $32.52^{+4.81}_{-4.66}$ \\
PG 2349-014 & $3.04^{+0.22}_{-0.24}$ & $22.51 \pm 6.21$ & $1.72^{+2.35}_{-1.32}$ & $55.90^{+11.00}_{-10.80}$ & $72.42^{+27.38}_{-20.66}$ \\
\hline
\end{tabular*}
{\raggedright\tablecomments{Column (1): object names. Column (2)--(4): disk size at 5100\AA\ computed from the SSD prediction, ICCF, and \texttt{TILDE} inference. Column (5): reverberation-mapped H$\beta$ lag reported by \cite{Woo2024}. Column (6): centroid line lag, $\tau_{\mathrm{line}, b} = (4/3)\tau^{(s)}_{\mathrm{line}, b}$, of the band with the largest H$\beta$ emission-line fraction. All quantities are in rest-frame}}
\label{results_table}
\end{table*}

The Stage~1 log posterior, $\ln \mathcal P$, combines the photometric likelihood with spectroscopic regularization terms derived from the synthesis of
\S\ref{sec:spec} and a set of physically motivated priors, $\ln \pi (\boldsymbol\theta)$ (detailed in Table~\ref{tab:tilde_priors})
\begin{equation}
\ln\mathcal{P}(\boldsymbol{\theta}\,|\,\mathrm{data})
   \;=\; \ln\mathcal{L}_{\mathrm{phot}}(\boldsymbol{\theta})
   \;+\; \ln\mathcal{S}(\boldsymbol{\theta})
   \;+\; \ln\pi(\boldsymbol{\theta})
\label{eq:logpost}
\end{equation}
with the spectroscopic term
\begin{equation}
\begin{aligned}
\ln\mathcal{S}(\boldsymbol{\theta})
={}& \sum_{b\in\mathcal{B}}
   \Bigl[
      \ln\pi_{\xi,b}(\xi_b)
      + \ln\pi_{F\mathrm{c},b}(F_{\mathrm{static},b}) \\
&\quad
      + \ln\pi_{F\mathrm{cont},b}
        (\bar F_{\mathrm{cont},b}\,|\,\boldsymbol{\theta})
   \Bigr] \\
&\quad
   + \ln\pi_{R_{F_{\!\mathrm{var}}}}(\boldsymbol{\theta})
\end{aligned}
\label{eq:logspec}
\end{equation}
The first two sums are Gaussian pulls toward $(\xi_b^{\rm spec},\sigma_{\xi,b})$ and $(F_{\mathrm{static},b}^{\rm spec},\sigma_{\mathrm{static},b})$; the third is a Gaussian in $\bar F_{\mathrm{cont},b}$ whose variance is inflated by the stationary variance of $c_b$ implied by the current $\boldsymbol{\theta}$, so the spectrophotometric continuum anchor becomes less restrictive when the model predicts large fractional disk variability. The soft log-normal penalty is given on the ratio of BLR to continuum fractional variability amplitudes, 
\begin{equation}
R_{F_{\rm var}, b} \equiv \frac{F_{\rm var, line}}{F_{\rm var, cont}} = \alpha \, \sqrt{\frac{\mathrm{Var}(l_b)}{\mathrm{Var}(c_b)}}
\end{equation}
where variances are analytically evaluated from the stationary moments of $c_b$ and $l_b$ at the sampled $\boldsymbol{\theta}$, 
\begin{equation}
\begin{aligned}
\mathrm{Var}(c) = \iint \Psi_{\rm cont}(\tau)\,\Psi_{\rm cont}(\tau')\,C_{\rm DRW}(\tau - \tau')\,d\tau \,d\tau' \\
\mathrm{Var}(l) = \iint \Psi_{\rm line}(\tau)\,\Psi_{\rm line}(\tau')\,C_{\rm DRW}(\tau - \tau')\,d\tau \,d\tau' 
\end{aligned}
\end{equation}
and acts as a physical regularizer when the photometric likelihood is locally degenerate between $(\xi_b,\alpha)$ and $(\bar F_{\mathrm{cont},b})$.

\begin{figure*}[ht!]
\centering
\includegraphics[width=\textwidth]{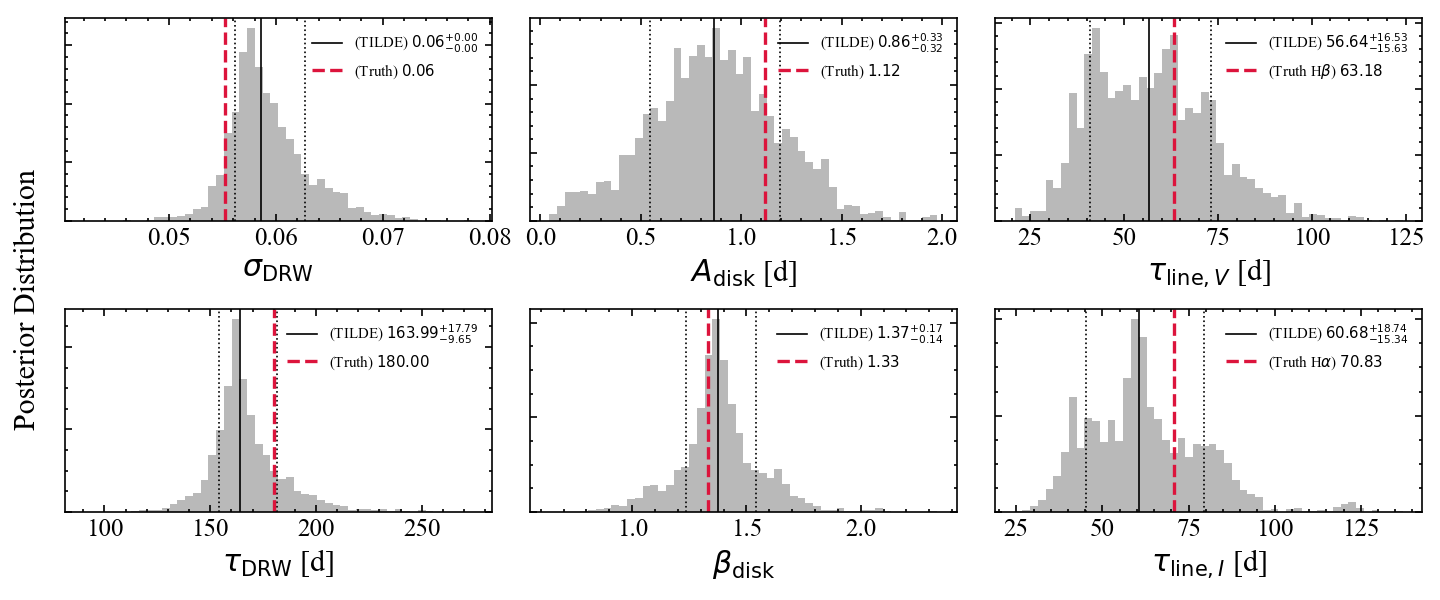}
\caption{
Posterior distributions (gray histogram) of representative \texttt{TILDE} parameters inferred from the fiducial \texttt{STiSS} target described in Section~\ref{subsec:example_decomp}. The parameters shown include the DRW driver parameters ($\sigma_{\mathrm{DRW}}$, $\tau_{\mathrm{DRW}}$), the disk-lag parameters ($A_{\mathrm{disk}}$, $\beta_{\mathrm{disk}}$), and the centroid BLR lags, $\tau_{\mathrm{line}, b} = (4/3)\tau^{(s)}_{\mathrm{line}, b}$, in the $V$ and $I$ bands. Solid black vertical lines indicate the posterior medians, dotted black vertical lines mark the central $1\sigma$ credible intervals, and red dashed vertical lines show the input truth values. All parameters are in the observed frame.
\label{fig:mock_verify}}
\end{figure*}

\begin{figure*}[ht!]
\centering
\includegraphics[width=\textwidth]{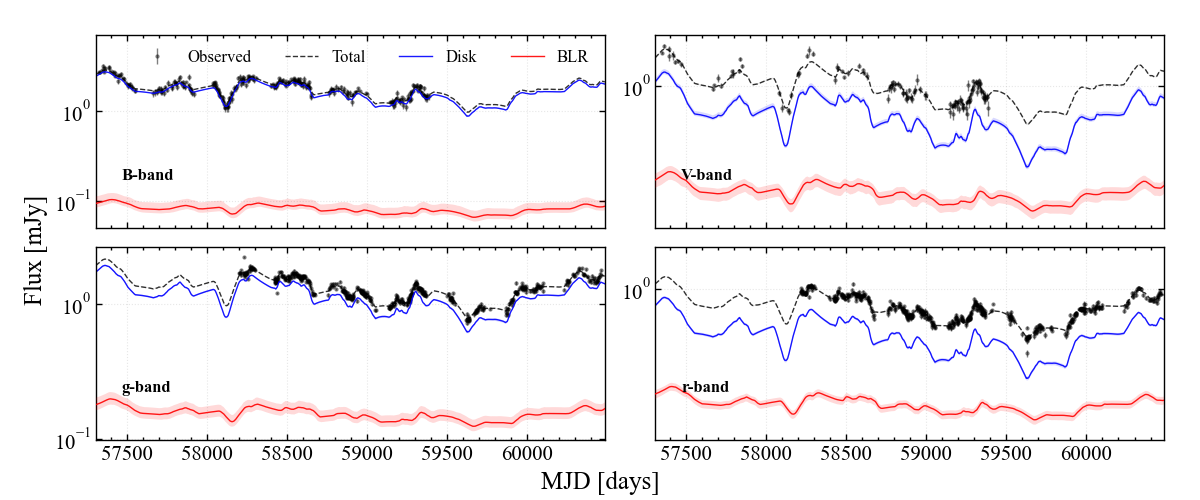}
\caption{
Example \texttt{TILDE} decomposition for the SAMP target J1026+523. The observed light curves (black points) are shown together with the best-fit total model (black line) and the decomposed disk (blue) and BLR (red) components. The $i$-band is excluded because the available spectrum does not sufficiently overlap the filter trasmission, as shown in Figure~\ref{fig:samp_spectral_decomposition}, to provide reliable spectroscopic constraints.
}
\label{fig:tilde_example_j1026}
\end{figure*}

\begin{figure*}[ht!]
\centering
\includegraphics[width=\textwidth]{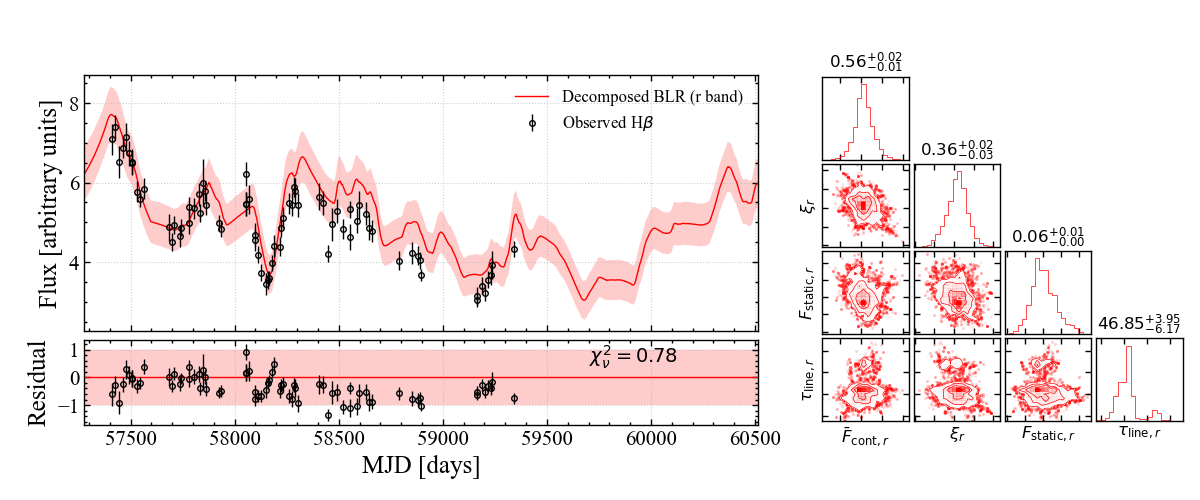}
\caption{Validation of the \texttt{TILDE}-decomposed BLR component for the SAMP target J1026+523. The left panel compares the reconstructed $r$-band BLR light curve (red solid line) with the independently observed spectroscopic H$\beta$ light curve (black points). The $r$ band is selected because it contains the largest H$\beta$ emission-line contribution for this target. The right panel shows the posterior distributions of the relevant $r$-band decomposition parameters in the observed frame. Note that $\tau_{\mathrm{line },r} = (4/3)\tau^{(s)}_{\mathrm{line },r}$.}
\label{fig:tilde_hbeta_validation_j1026}
\end{figure*}
\begin{figure}[!t]
\centering
\includegraphics[width=0.95\columnwidth]{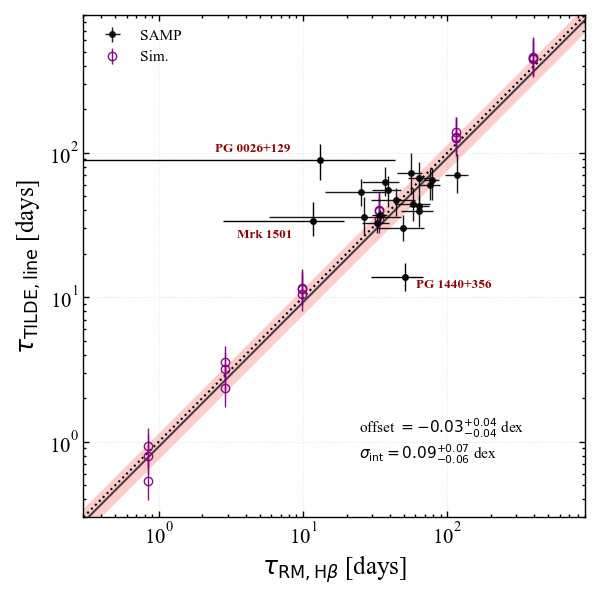}
\caption{Comparison between reference H$\beta$ lags and \texttt{TILDE}-inferred effective line lags, $\tau_{\mathrm{line}, b} = (4/3)\tau^{(s)}_{\mathrm{line}, b}$. Reference lag of the SAMP targets (red) and mock simulations (purple) are the spectroscopic H$\beta$ lag and known input lag, respectively. For each target, the \texttt{TILDE} lag is taken from the filter with the largest H$\beta$ emission-line fraction. Sample offset (gray solid line) and intrinsic scatter (gray shade) are calculated and plotted for the SAMP results. The dotted line shows the one-to-one relation. }
\label{fig:tilde_lrm_lag_comparison}
\end{figure}

\begin{figure*}[!t]
\centering
\includegraphics[width=\textwidth]{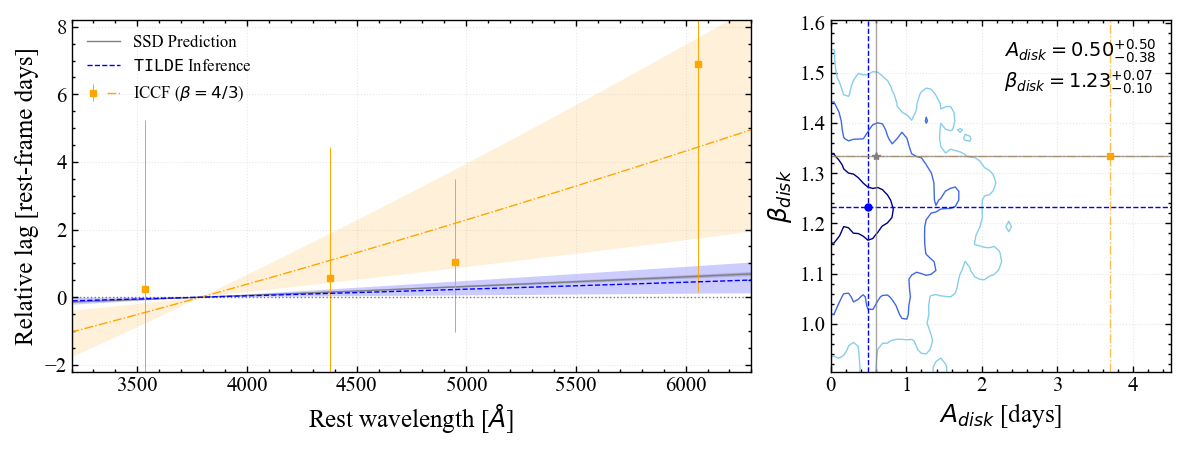}
\caption{
Left: lag spectrum of J1026+523 computed from the SSD prediction (solid gray line), ICCF lag measurement (orange squares) and disk fitting (orange dot-dashed line) reported by \cite{Mandal2025}, and TILDE inference ($\tau_{\mathrm{cont},b} = 1.5 \,\tau^{(s)}_{\mathrm{cont},b}$; blue dashed line). Right: two-dimensional posterior distribution from TILDE for $A_{\rm disk}$ and $\beta_{\rm disk}$, with the best-fit values shown as blue dashed lines.
}
\label{fig:tilde_disk_lags_example}
\end{figure*}
\begin{figure}[!t]
\centering
\includegraphics[width=0.95\columnwidth]{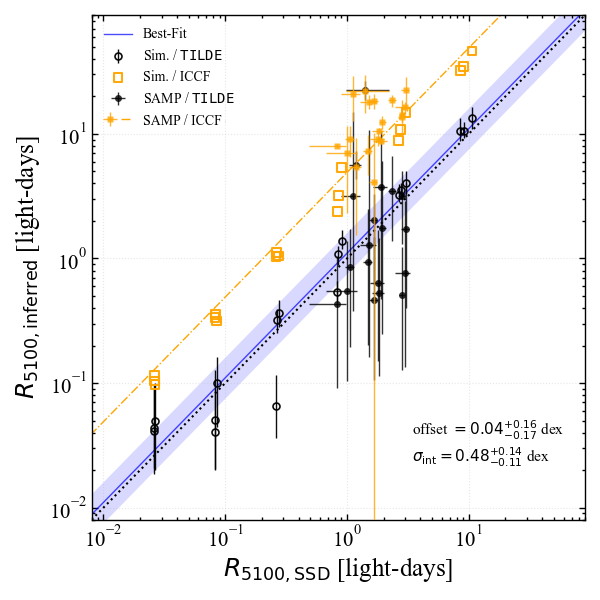}
\caption{Comparison between the SSD-predicted disk size, $R_{5100,{\rm SSD}}$, and the inferred disk size, $R_{5100,{\rm inferred}}$, based on ICCF (orange) and \texttt{TILDE} (black) for \texttt{STiSS} (open symbols) and SAMP (filled symbols) samples. The \texttt{TILDE}-based sample offset (blue line) and intrinsic scatter (blue shading), calculated for the SAMP sample, are plotted along with the ICCF-based sample offset (orange dot-dashed line).}

\label{fig:tilde_disk_size_comparison}
\end{figure}

\subsubsection{Optimization and Sampling}
\label{sec:stage1-opt}

The photometric likelihood of Eq.~\eqref{eq:likelihood} is a function of a $(6+4|\mathcal{B}|)$-dimensional vector through a covariance matrix that encodes multiple shifted DRW kernels; its posterior landscape is typically multi-modal in $A_{\mathrm{disk}}$ and in the $\tau^{(s)}_{\mathrm{line},b}$ block, and exhibits a family of approximate degeneracies among $(\xi_b,\alpha,\bar F_{\mathrm{cont},b})$ that the spectral penalties are designed to break. To locate a defensible MAP point under these conditions, we perform a multi-start L-BFGS-B optimization of $-\ln\mathcal{P}(\boldsymbol{\theta})$ with ten randomized restarts. Starting points are drawn parameter-by-parameter around a data-driven seed: $A_{\mathrm{disk}}$ receives a broad uniform perturbation in log space to sample distinct lag wells, each $\tau^{(s)}_{\mathrm{line},b}$ receives a moderate Gaussian perturbation in log space, and the remaining parameters receive tight log-space perturbations. The best-fit restart, scored by the final value of $-\ln\mathcal{P}$, is retained as the MAP estimate $\boldsymbol{\theta}_{\mathrm{MAP}}$.

Starting from $\boldsymbol{\theta}_{\mathrm{MAP}}$, we run an affine-invariant ensemble sampler using \texttt{EMCEE} \citep{ForemanMackey2013} with 64 walkers for 400 steps, discarding the first 150 as burn-in. Walker initial positions are drawn with parameter-specific widths: a $\sigma=0.05$ Gaussian ball in log space for $A_{\mathrm{disk}}$, a $\sigma=0.5$ Gaussian ball in log space for each $\tau^{(s)}_{\mathrm{line},b}$, and a tight $\sigma\sim10^{-4}$ fractional ball for the remaining parameters. We explicitly treat the resulting chain as a \emph{local posterior characterization} of $\mathcal{P}(\boldsymbol{\theta} \,|\,\mathrm{data})$ in the neighborhood of the MAP mode, sufficient to propagate correlated parameter uncertainties into Stage~2 but not as a global multi-modal posterior exploration. Identification of secondary modes is therefore delegated to the multi-start MAP search of \S\ref{sec:stage1-opt}, which is explicitly designed for that role. We monitor integrated autocorrelation times for each parameter and discard chains whose autocorrelation is incompatible with the nominal chain length.

Regarding parameter transforms during computation, most positive-definite parameters are sampled in logarithmic space; the Jacobian of this reparameterization is folded into the prior term where it would otherwise distort the marginal (for example, the explicit $+\ln\xi_b$ correction). $F_{\mathrm{static},b}$ is sampled through a softplus transform to enforce non-negativity without a hard boundary. These transforms are implementation conveniences; they do not alter the posterior and we do not refer to them elsewhere in the paper.

\subsection{Stage 2: Gaussian-Process Reconstruction}
\label{sec:stage2}

Given a posterior sample $\boldsymbol{\theta}_s$ drawn from the Stage~1 chain, we condition the latent DRW driver $X(t)$ on the data by standard GP regression. Let $\mathbf{K}_{x}$ be the DRW prior covariance evaluated on a dense reconstruction grid $\{t_j\}$, and $\mathbf{K}_{\times}$ the cross-covariance between the grid points and the observed data positions.
The conditional mean and covariance of $x$ on the grid are
\begin{align}
\mathbf{m}_{x} &\;=\; \mathbf{K}_{\times}\,\mathbf{C}^{-1}(\mathbf{Y}-\boldsymbol{\mu}),
\label{eq:gp-mean-cov}\\
\boldsymbol{\Sigma}_{x} &\;=\; \mathbf{K}_{x} \;-\;
   \mathbf{K}_{\times}\,\mathbf{C}^{-1}\,\mathbf{K}_{\times}^{\mathsf{T}}
\label{eq:gp-cov}
\end{align}
with $\mathbf{C}$ and $\boldsymbol{\mu}$ evaluated at $\boldsymbol{\theta}_s$. We then reconstruct the latent disk-continuum component in band $b$ by evaluating $\mathbf{m}_x$ on the grid shifted by $\tau_{\mathrm{cont},b}$ (with the M{=}2 two-tap weighting of Eq.~\eqref{eq:disk-k2}), and the latent BLR component by evaluating $\mathbf{m}_x$ on the grid shifted by $\tau_{\mathrm{line},b}$. Recombining with Eq.~\eqref{eq:forward} yields per-band reconstructions of $F_b(t)$, $\bar F_{\mathrm{cont},b}[1+c_b(t)]$, and $\xi_b\,\bar F_{\mathrm{cont},b}[1+\alpha\,l_b(t)]$, each with posterior bands obtained by repeating the reconstruction across several hundred Stage~1 samples.

\section{Results}
\label{Results}

We apply \texttt{TILDE} to both the synthetic data from \texttt{STiSS} and the SAMP light curves. First, we simulate a total of 18 AGN spanning $\log \left( L_{5100} / {\rm erg~s^{-1}} \right) = \{41, 42, 43, 44, 45, 46\}$ and $\lambda_{\rm Edd} = \{0.01, 0.1, 1.0\}$ located at $z=0.2$ (\S\ref{sec:mock_data}). Second, we use the photometric and spectroscopic light curves from the SAMP for a sample of 19 AGN (\S\ref{subsec:samp_data_details}). 

We first present example \texttt{TILDE} decompositions for a \texttt{STiSS} and a SAMP target (\S\ref{subsec:example_decomp}). We then examine the two main reconstructed components separately for the full sample: the decomposed BLR light curves and effective BLR lags (\S\ref{subsec:decomposed_blr}), followed by the decomposed disk light curves and the resulting impact on disk-size measurements (\S\ref{subsec:decomposed_disk}), summarized in Table ~\ref{results_table}.

\subsection{Example decompositions}
\label{subsec:example_decomp}

We first show an example \texttt{TILDE} run on \texttt{STiSS} for an AGN with $\log \left( L_{5100}/{\rm erg~s^{-1}} \right)=44$ and $\lambda_{\rm Edd}=0.1$. \texttt{TILDE} is applied to the extracted $BVRI$ light curves together with a single-epoch spectrum randomly selected from the corresponding spectral time series; then the inferred parameters are compared with the known input values. Despite the approximations of the \texttt{TILDE} forward model, the recovered posterior distributions are consistent with the input parameters for the driver, disk response, and BLR lags, as shown in Figure~\ref{fig:mock_verify}. The amplitude and damping timescale of the DRW variability are recovered as $\sigma_{\rm DRW}=0.06^{+0.00}_{-0.00}$ and $\tau_{\rm DRW}=163.99^{+17.79}_{-9.65}$ days, consistent with the input values of $0.06$ and $180$ days (150 days in rest frame). The disk-lag parameters are also recovered within the posterior uncertainties, with $A_{\rm disk}=0.86^{+0.33}_{-0.32}$ days compared to the input value of $1.12$ days, and $\beta_{\rm disk}=1.37^{+0.17}_{-0.14}$ compared to the input value of $1.33$.

Since we set a source redshift of $z=0.2$ in the simulation, H$\beta$ and H$\alpha$ emission lines fall primarily within the $V$ and $I$ bands, respectively. The inferred effective BLR lags, accounted for the first-moment correction factor, in the line-dominated bands are likewise consistent with the true emission-line delays. In the $V$ band, where H$\beta$ contributes strongly to the broad-band flux, \texttt{TILDE} recovers $\tau_{{\rm line},V}=56.64^{+16.53}_{-15.63}$ days, compared to the true H$\beta$ lag of $63.18$ days. In the $I$ band, where H$\alpha$ dominates the line contribution, \texttt{TILDE} recovers $\tau_{{\rm line},I}=60.68^{+18.74}_{-15.34}$ days, compared to the true H$\alpha$ lag of $70.83$ days. These results demonstrate that the simplified covariance-based model of \texttt{TILDE} is sufficiently flexible to separate the prompt disk response from the more delayed BLR contribution in broad-band photometry.

We next show an example decomposition for the SAMP target J1026+523. To avoid biased priors caused by inadequate spectral coverage, we excluded filters whose transmission curves overlapped the observed spectrum by less than 80\% (an example of the excluded $i$ band for J1026+523 is shown in Figure~\ref{fig:tilde_example_j1026}). \texttt{TILDE} adequately fits the observed multi-band light curves and separates the broad-band variability into disk and BLR components. To empirically validate the decomposition, we compare the reconstructed BLR light curve in the band containing the H$\beta$ emission line with the independently observed spectroscopic H$\beta$ light curve. Although the selected band may also contain other BLR-related components, the dominance of broad H$\beta$ emission allows this comparison to serve as an empirical test. As shown in Figure~\ref{fig:tilde_hbeta_validation_j1026}, the close agreement between the reconstructed and observed light curves indicates that \texttt{TILDE} can recover delayed BLR variability from broad-band photometry.

\begin{figure*}[ht!]
\centering
\includegraphics[width=\textwidth]{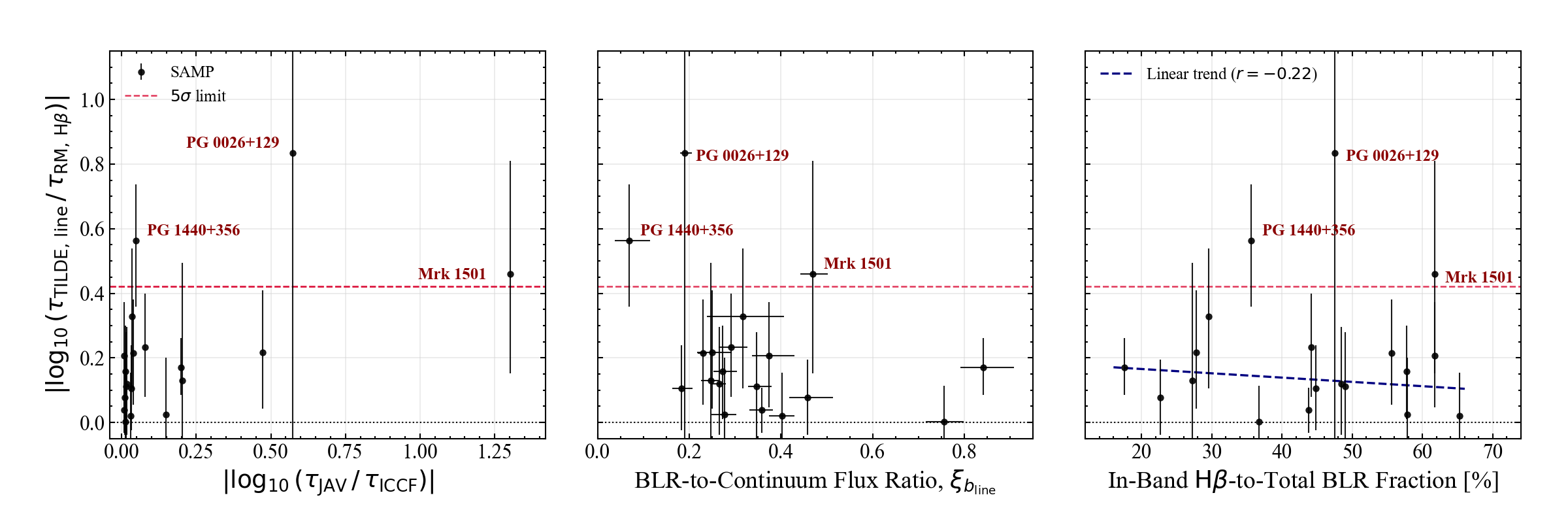}
\caption{Discrepancy between the reference H$\beta$ lags and the \texttt{TILDE}-inferred effective BLR lags, shown as a function of the discrepancy between the reported \texttt{JAVELIN} and ICCF lags (left), the BLR-to-continuum flux ratio (center), and the in-band H$\beta$-to-total BLR flux fraction (right). The 5$\sigma$ limit, shown as the horizontal dashed line, is computed from mean offset and intrinsic scatter.The Pearson correlation coefficient is calculated after excluding the three outliers.}
\label{fig:outlier_analysis}
\end{figure*}

\subsection{Decomposed BLR light curves}
\label{subsec:decomposed_blr}

We present the full-sample comparison of BLR lags inferred from the decomposed light curves with the known input lags for \texttt{STiSS} and the observed H$\beta$ lags for SAMP (Figure~\ref{fig:tilde_lrm_lag_comparison}). For the \texttt{STiSS} sample, the \texttt{TILDE}-inferred lags show negligible discrepancy from the input lags over the simulated range of luminosity and Eddington ratio, demonstrating that the lag recovery remains accurate across the full sample. For the SAMP sample, the mean offset between the RM-based H$\beta$ lags and the \texttt{TILDE}-inferred effective BLR lags is $-0.03^{+0.04}_{-0.04}$ dex, indicating no significant systematic discrepancy between the two measurements. The inferred intrinsic scatter is $0.09^{+0.07}_{-0.06}$ dex, suggesting that \texttt{TILDE} successfully traces the delayed BLR component from broad-band photometry while still leaving object-to-object differences beyond the mean agreement. This scatter is expected because the \texttt{TILDE}-inferred lag is an effective broad-band BLR-related lag, weighted by all reverberating components within the selected filter, rather than the lag of a unique physical emission component. The interpretation and limitations of this effective-lag measurement are discussed further in \S\ref{sec:discussion_blr_size} and Section~\ref{sec:discussion_limitations_future}, respectively.

\subsection{\texttt{TILDE} Disk Mapping}
\label{subsec:decomposed_disk}

With the successful decomposition of the disk and BLR components, we quantify the systematic effect of the BLR contamination on the inter-band lag using the 19 SAMP AGN. By comparing the ICCF (Interpolated Cross-Correlation Function) lags of the broad-band with the largest BLR contribution with the corresponding \texttt{TILDE}-inferred lags, we find that BLR contamination increases the inter-band lags by $0.69\pm0.56$ dex. Consequently, this effect will lead to an overestimation of the disk size. For example, we demonstrate that the lag spectrum inferred from the \texttt{TILDE} is consistent with the SSD expectation for an AGN in Figure~\ref{fig:tilde_disk_lags_example}. \texttt{TILDE} constrains disk properties with $A_{\rm disk}=0.50^{+0.50}_{-0.38}$ days and $\beta_{\rm disk}=1.23^{+0.07}_{-0.10}$, consistent with the SSD expectation.

We compare the SSD predictions with the disk sizes inferred from the broad-band lags and from \texttt{TILDE} in Figure~\ref{fig:tilde_disk_size_comparison}. We measure the response-weighted effective disk size at the rest-frame wavelength 5100\AA\ using
\begin{equation}
R_{5100} = c A_{\rm disk}
\left(\frac{5100\,\text{\AA}}{\lambda_{\rm ref}}\right)^{\beta_{\rm disk}}
\end{equation}
where $\lambda_{ref}$ is the effective wavelength of the $g$-band filter. For the \texttt{STiSS} sample, applying ICCF directly to the broad-band light curves produces systematically larger disk sizes than the input sizes of the SSD disk used in the \texttt{STiSS}. However, after applying \texttt{TILDE}, the recovered disk sizes agree with the input SSD sizes, demonstrating that \texttt{TILDE} can successfully derive the disk lag by removing the delayed BLR contribution. Similar behavior is seen in the SAMP sample. The disk sizes measured from the broad-band light curves are larger than the SSD prediction by a factor of $\sim 4.9$ as reported by \citet{Mandal2025}. In contrast, the disk sizes after removing the BLR contribution using \texttt{TILDE} become consistent with the SSD prediction with an average ratio between the two $R_{\texttt{TILDE}}/R_{\rm SSD}=1.12^{+0.50}_{-0.37}$ with an intrinsic scatter of $\sigma_{\rm int} = 0.48^{+0.14}_{-0.11}$.

\section{Discussion}
\label{sec:discussion}

\subsection{Interpretation of BLR and Disk Lags}
\label{sec:discussion_blr_size}

The \texttt{TILDE} recovers BLR light curves and lags without relying on spectroscopic monitoring data. As shown in Figure~\ref{fig:tilde_lrm_lag_comparison}, the \texttt{TILDE}-inferred BLR lag is consistent with the reverberation-mapped H$\beta$ lag with a modest intrinsic scatter, suggesting that \texttt{TILDE} can be widely used for time-domain photometry data. Since BLR lags can be inferred without expensive spectroscopic campaigns, BLR lags can be estimated for a substantially larger sample of AGN than the current spectroscopic RM AGN.

However, the \texttt{TILDE}-inferred BLR lags should be interpreted with caution in cases where either the underlying RM lag measurement or the broad-band decomposition is poorly constrained, as illustrated by the three outliers in Figure~\ref{fig:tilde_lrm_lag_comparison}. Mrk~1501 shows the largest discrepancy between the reported DRW-based \texttt{JAVELIN} \citep[][]{Zu2011} and interpolation-based ICCF lags (Figure~\ref{fig:outlier_analysis}), which \citet{Woo2024} attributed this discrepancy to aliasing caused by sparse sampling and/or possible mismatch between the intrinsic variability and the assumed DRW process \citep{Kozlowski2017}. Related stochastic-modeling analyses of photometric RM have likewise shown that reliable lag recovery can be limited by weak line contributions and finite photometric precision \citep{Zu2016}. Because \texttt{TILDE} is fitted to the same finite and irregularly sampled light curves and likewise adopts a DRW driver, these sources of uncertainty can also degrade the photometric decomposition. PG~0026+129 provides a related example: its ICCF lag was reported to be unresolved by \citet{Woo2024}, and likewise deviates from the general relation. Finally, a small BLR-to-continuum flux ratio can render the delayed BLR signal difficult to distinguish from measurement noise. PG~1440+356, another outlier, has the lowest inferred ratio in the SAMP sample, $\xi = 0.19^{+0.04}_{-0.04}$, consistent with this interpretation.

Furthermore, note that the \texttt{TILDE}-inferred BLR lag can be slightly different from the lag of any single broad emission line. By construction, the variability ratio $\xi_{b}$ inherits contributions from the combination of broad emission lines, the Fe~{\sc ii} pseudo-continuum, and the potential diffuse continuum emission that fall within the broad-band filter, all of which can reverberate on slightly distinct timescales. Recent velocity-resolved RM results further suggest that even a single broad emission line can arise from multiple reverberating BLR components with distinct delay distributions \citep{LongDexter2025}. The inferred lag is therefore shifted away from the H$\beta$ lag as the relative contribution of H$\beta$ to the in-band BLR flux decreases, reflected in Figure~\ref{fig:outlier_analysis}. Hence, the recovered lag is best understood as an effective BLR lag, weighted by the variable components transmitted through the selected filter, rather than as the lag of a single broad emission line.

The removal of the delayed BLR contribution also brings the disk sizes inferred from the SAMP light curves into substantially better agreement with the SSD prediction. This result is consistent with the microlensing analysis of \citet{FianChelouche2023}, who showed that BLR contribution can account for the apparently enlarged continuum-emitting regions inferred when the observed broad-band flux is interpreted as arising solely from an SSD. In both cases, the tension with the SSD is reduced once the non-disk BLR contribution is accounted for, suggesting that at least part of the disk-size discrepancy originates from the interpretation of the observed signal rather than from a failure of the disk model itself. However, the intrinsic scatter, $\sigma_{\rm int}=0.48^{+0.14}_{-0.11}$ dex, may imply deviation from the standard thin-disk structure, potentially including changes in the accretion-flow geometry at high accretion rates, such as the onset of a geometrically thicker, advective slim-disk regime \citep[e.g.,][]{Abramowicz1988, Donnan2023}. By separating prompt disk variability from delayed BLR contamination using only broad-band photometry, \texttt{TILDE} can enable a less biased investigation of such deviations across the much larger AGN samples accessible to future time-domain surveys.

\subsection{Limitations}
\label{sec:discussion_limitations_future}

Several limitations should be considered when interpreting our results. First, the \texttt{TILDE} inference depends on the adopted parameterization of the disk and BLR responses. The adopted two-tap kernels may not be sufficient to accurately represent the true disk and BLR transfer functions, and the resulting model mismatch may be absorbed into the inferred parameters unless higher-order approximations are considered. More generally, tests of reverberation-mapping inference have shown that simplified response models can reproduce observed variability while yielding an incomplete physical interpretation when the assumed BLR structure differs from the underlying system \citep{Mangham2019}. Second, \texttt{TILDE} relies on the single-epoch spectroscopic decomposition described in \S\ref{sec:spec} to separate static and variable contributions, assuming that the line-to-continuum flux ratio and static component remain effectively constant over the monitoring baseline. Third, degeneracies among the disk lag, BLR lag, band-dependent flux ratios, and photometric jitter can increase the marginalized parameter uncertainties. The MAP optimization and local \texttt{EMCEE} sampling described in \S\ref{sec:stage1-opt} are designed to mitigate, but not eliminate, the impact of these correlated posterior directions. Finally, the current \texttt{TILDE} implementation requires substantial computational resources, with each SAMP AGN requiring approximately $\sim$1 hr of inference time. Scaling the method to large surveys will therefore require computational optimization, potentially through simulation-trained, machine-learning-assisted inference, as in recent latent stochastic differential-equation approaches \citep[e.g.,][]{Fagin2025}, or by applying it to a carefully selected representative sample.

The validation is also limited by the physical scope of \texttt{STiSS}, which is intentionally designed as a controlled test-data generator rather than a comprehensive model of all AGN emission processes. In particular, diffuse Balmer-continuum and dust emission should be incorporated into \texttt{STiSS} in future work to assess their impact on the decomposition before applying \texttt{TILDE} to surveys with finer or broader wavelength coverage, where these additional components may contribute variability on distinct timescales. More comprehensive AGN forward simulations have begun to treat diffuse-continuum and dusty-torus effects alongside disk and BLR variability \citep[e.g.,][]{Best2025}; incorporating such effects into \texttt{STiSS} would provide a more stringent stress test of \texttt{TILDE}.

\section{Summary}
\label{sec:summary}

We present \texttt{TILDE}, a two-stage Bayesian framework for decomposing multi-band AGN broad-band light curves into accretion-disk and BLR components, and validate it using mock synthetic datasets and the SAMP AGN sample. Our main results are summarized as:

\begin{enumerate}
    \item We develop a forward model in which each broad-band light curve is represented as the sum of a static component, a variable disk-continuum component, and a delayed BLR component. Decomposition of single-epoch spectrum is used to regularize the line-to-continuum flux ratios and static flux contributions in each band.

    \item By applying \texttt{TILDE} to the synthetic data from SED simulations, we verify that \texttt{TILDE} successfully recovers the input DRW driver, disk lags, and BLR lags by decomposing the underlying disk and BLR light curves. This demonstrates that the simplified covariance-based model can separate prompt disk variability from delayed BLR-related variability in multi-broad-band photometry. By applying \texttt{TILDE} to the SAMP sample, we find \texttt{TILDE} reproduces the observed spectroscopic H$\beta$ light curves. The spectroscopic H$\beta$ lags and \texttt{TILDE}-inferred BLR lags are consistent with an intrinsic scatter of $\sigma_{\rm int}=0.09^{+0.07}_{-0.06}$ dex.

    \item After removing BLR contamination, the \texttt{TILDE} inferred disk sizes are more consistent with the standard thin-disk expectation, with an average ratio of the two $R_{\rm inferred}/R_{\rm SSD}=1.12^{+0.50}_{-0.37}$, suggesting that delayed BLR contamination can substantially bias broad-band continuum reverberation measurements and may partly resolve the oversized disk problem. 
\end{enumerate}

Future detailed studies can be performed by applying \texttt{TILDE} to large photometric AGN surveys, i.e., ZTF and LSST, where multi-band light curves can extend reverberation analyses to samples far larger than those accessible to spectroscopic campaigns. In this regime, \texttt{TILDE} enables accretion-disk studies and photometric BLR-size inference for AGN that are difficult to monitor spectroscopically.

\begin{acknowledgments}

We thank Hyungsuk Tak for insightful comments that improved the clarity of the statistical methodology. This work has been supported by the Basic Science Research Program through the National Research Foundation of Korean Government (RS-2026-25487266).
\end{acknowledgments}

\bibliography{references}{}
\bibliographystyle{aasjournalv7}

\appendix
\section{Prior-to-Posterior Updates of the Effective BLR Lag}
\label{app:prior_posterior_lags}

\begin{figure*}[ht!]
\centering
\includegraphics[width=\textwidth]{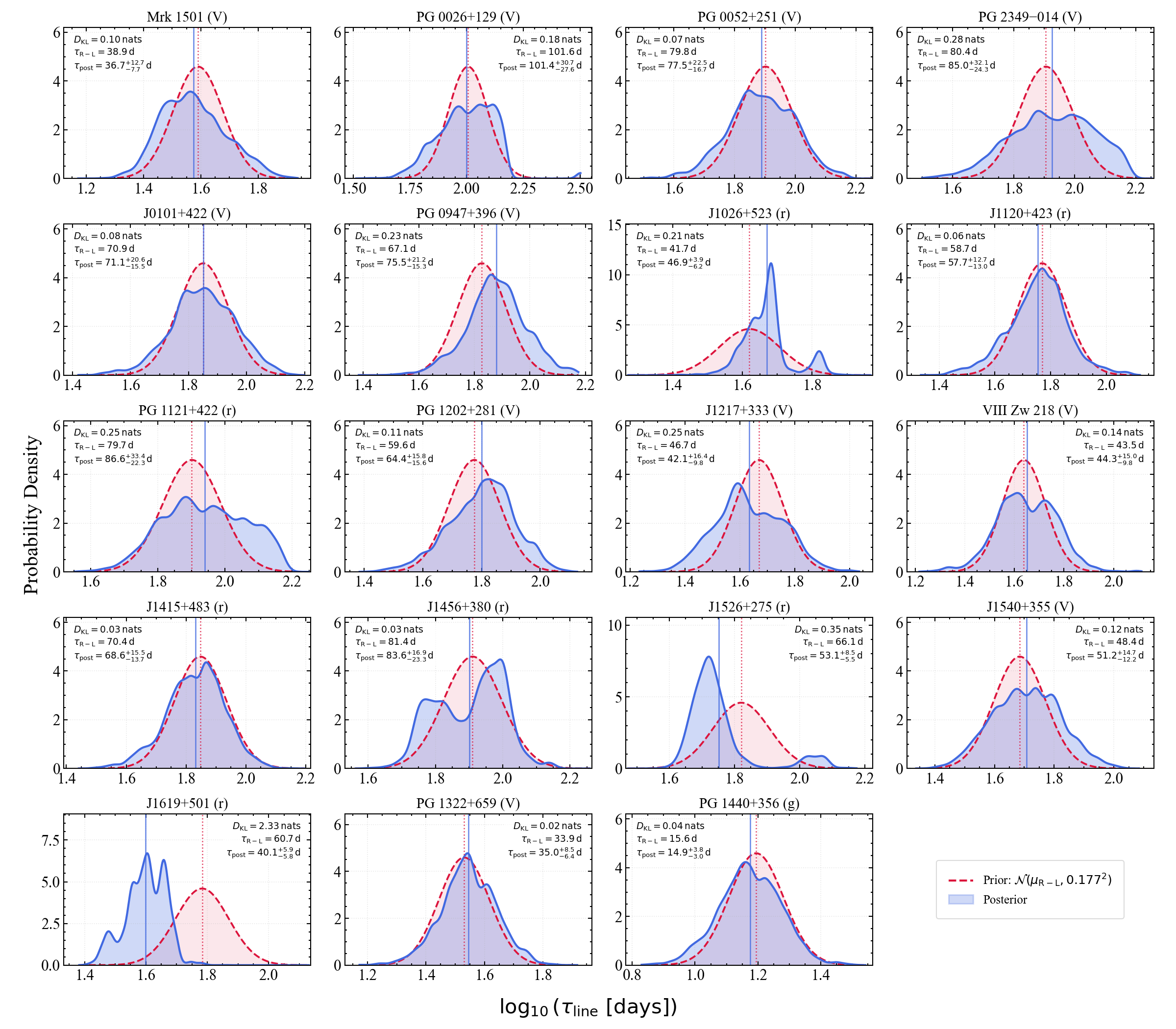}
\caption{Prior-to-posterior comparison for the effective BLR lag in the SAMP band with the most H$\beta$ emission. The red dashed curve shows the adopted prior and the blue shaded curve shows the marginalized Stage~1 posterior. Red dotted and blue solid vertical lines mark the prior mean and posterior median, respectively. Each panel lists the KL divergence, $D_{\rm KL}(\mathcal{P}\Vert\pi)$, evaluated in $\log_{10}(\tau_{\rm line}/{\rm day})$ and reported in nats.}
\label{fig:KL_divergence}
\end{figure*}

The effective broad-band BLR lags inferred by \texttt{TILDE} are regularized by the H$\beta$ $RL$ relation described in \S\ref{sec:spec}. We therefore examine whether the inferred lag posteriors remain tied to this population-level expectation or can depart from it when the joint multi-band light-curve model favors a different solution. This test is particularly relevant because the sampled parameter $\tau_{{\rm line},b}$ represents an effective BLR-related lag in a broad-band filter, whereas the adopted $RL$ relation predicts the lag of the H$\beta$ emission line.

For each SAMP target, we select the band $b_{\rm H\beta}$ whose effective wavelength is closest to the observed H$\beta$ wavelength. We then compare the marginalized Stage~1 posterior of $\tau_{{\rm line},b_{\rm H\beta}}$ with the corresponding $RL$ prior. Both distributions are expressed in the same time frame as the sampled lag parameter. Figure~\ref{fig:KL_divergence} presents these comparisons for the 19 SAMP targets.

We quantify the prior-to-posterior update using the Kullback--Leibler (KL) divergence,
\begin{equation}
D_{\rm KL}\left(\mathcal{P}\Vert\pi\right) = 
\int\mathcal{P}(x)\ln\left[\frac{\mathcal{P}(x)}{\pi(x)}\right] dx
\label{eq:kl_lag}
\end{equation}
where $x=\log_{10}(\tau_{{\rm line},b_{\rm H\beta}}/{\rm day})$, $\pi(x)$ is the adopted $RL$ prior, and $\mathcal{P}(x)$ is the marginalized posterior. We evaluate $\mathcal{P}(x)$ using a Gaussian kernel-density estimate of the Stage~1 samples, allowing the calculation to retain posterior asymmetry or multimodality. The KL divergence is reported in nats and measures the extent to which the full inference updates the adopted lag prior.

A close correspondence between a posterior and its prior is expected for targets whose effective broad-band BLR lag is consistent with the H$\beta$ relation. Thus, the purpose of this diagnostic is not to require a large prior-to-posterior shift for every target. Instead, it tests whether \texttt{TILDE} can respond to target-specific information in the multi-band light curves rather than mechanically reproducing the $RL$ regularization.

The resulting prior-to-posterior updates are strongly target dependent. Many targets remain broadly consistent with the $RL$ expectation, while others show appreciable shifts or changes in posterior shape. In particular, J1619+501 exhibits a substantial displacement from the prior mean, with $D_{\rm KL}=2.33$~nats and a posterior-mode shift of $2.13\sigma$ relative to the prior width. J1526+275 also shows a noticeable update, with $D_{\rm KL}=0.35$~nats and a shift of $0.79\sigma$. These cases demonstrate that the inferred effective BLR lag is not fixed by the $RL$ relation. Rather, the prior provides regularization for targets whose photometric reverberation signature is consistent with the population expectation, while allowing the joint light-curve model to favor a different lag when supported by the data. Therefore, the diversity of prior-to-posterior updates demonstrates that the effective BLR-lag posteriors are not uniformly inherited from the $RL$ prior.

\end{document}